\RequirePackage[2020-02-02]{latexrelease}
\documentclass[%
reprint,
amsmath,amssymb,
aps,
]{revtex4-2}
\usepackage{longtable}
\usepackage{graphicx}
\usepackage{dcolumn}
\usepackage{bm}

\begin{document}
	\preprint{APS/123-QED}
	\title{Identification of newly observed singly charmed baryons using relativistic flux tube model}
	\author{Pooja Jakhad$ ^{1*} $}
	\author{Juhi Oudichhya$ ^{1} $}
	\author{Keval Gandhi$ ^{2} $}
	\author{Ajay Kumar Rai$ ^{1} $}
	\affiliation{$ ^{1*} $Department of Physics, Sardar Vallabhbhai National Institute of Technology, Surat, Gujarat-395007, India}
	\affiliation{$ ^{2} $Department of Computer Science and Engineering, Institute of Advanced Research, Gandhinagar, Gujarat-382426, India}
	%
	%
	\date{\today}
	
\begin{abstract}
We calculate the mass spectra of $\Lambda_{c}$,  $\Xi_{c}$, $\Sigma_{c}$, $\Xi_{c}^{'}$, and $\Omega_{c}$ baryons in the framework of quark-diquark configuration using relativistic flux tube model.
The spin-dependent interactions are included in the j-j coupling scheme to find complete mass spectra.
We satisfactorily describe the known singly charmed baryons in quark-diquark configuration.
The possible spin-parity $ J^{P} $ quantum numbers are assigned to several experimentally observed states. 
Furthermore, some useful mass predictions are given for more excited states that are reasonably consistent with other model predictions for lower excited states. {From the obtained results the Regge trajectories for these singly charmed baryons are constructed in the ($J,M^{2}$) plane and the properties like linearity, parallelism and equidistant are verified.} Also, these predictions should be tested in future experiments.
\end{abstract}

	\maketitle
	
\section{Introduction}	
\begin{table}
	\caption{\label{tab:table1}
		Masses and $J^{P}$ values of the experimentally observed single-charmed baryons as specified in Particle Data Group \cite{pdg2022}. The status is listed as poor (*), only fair (**), very likely to certain (***), and certain (****) for existence.
	}
	\begin{ruledtabular}
		\begin{tabular}{lllllllll}
			State   & Mass (MeV) & $J^{P}$ & Status \\
			\colrule\noalign{\smallskip}
			$\Lambda_{c}^{+}$ &  2286.46 $\pm$ 0.14 & $\frac{1}{2}^{+}$ & ****\\
			$\Lambda_{c}(2595)^{+}$ & 2592.25 $\pm$ 0.28 & $\frac{1}{2}^{-}$ & ***\\
			$\Lambda_{c}(2625)^{+}$ & 2628.11 $\pm$0.19 & $\frac{3}{2}^{-}$ &  ***\\
			$\Lambda_{c}(2765)^{+}$/$\Sigma_{c}(2765)^{+}$ & 2766.6$  \pm $2.4 & $?^{?}$ & *\\
			$\Lambda_{c}(2860)^{+}$ & 2856.1 $ {}^{+2.3}_{-6.0} $ & $\frac{3}{2}^{+}$ & ***\\
			$\Lambda_{c}(2880)^{+}$ & 2881.63 $ \pm $0.24 & $\frac{5}{2}^{+}$ & ***\\
			$\Lambda_{c}(2940)^{+}$ & 2939.6 $ {}^{+1.3}_{-1.5} $ & $\frac{3}{2}^{-}$ & ***\\
			
			$\Sigma_{c}(2455)^{++}$ & $2453.97 \pm0.14  $ & $\frac{1}{2}^{+}$ & ****\\
			$\Sigma_{c}(2455)^{+}$ & $ 2452.65 {}^{+0.22}_{-0.16} $ & $\frac{1}{2}^{+}$ & ****\\
			$\Sigma_{c}(2455)^{0}$ & $ 2453.75 \pm0.14	 $ & $\frac{1}{2}^{+}$ & ****\\
			$\Sigma_{c}(2520)^{++}$ &$ 2518.41 {}^{+0.22}_{-0.18} $  & $\frac{3}{2}^{+}$ & ***\\
			$\Sigma_{c}(2520)^{+}$ &$ 2517.4 {}^{+0.7}_{-0.5} $  & $\frac{3}{2}^{+}$ & ***\\
			$\Sigma_{c}(2520)^{0}$ &$ 2518.48 \pm0.20 $  & $\frac{3}{2}^{+}$ & ***\\
			$\Sigma_{c}(2800)^{++}$ &$ 2801 {}^{+4}_{-6} $  & $?^{?}$ & ***\\
			$\Sigma_{c}(2800)^{+}$ &$2792 {}^{+14}_{-5}  $  & $?^{?}$ & ***\\
			$\Sigma_{c}(2800)^{0}$ &$ 2806 {}^{+5}_{-7} $  & $?^{?}$ & ***\\
			
			$\Xi_{c}^{+}$  & $ 2467.71 \pm0.23 $ & $\frac{1}{2}^{+}$ & ***\\
			$\Xi_{c}^{0}$  & $ 2470.44 \pm0.28 $ & $\frac{1}{2}^{+}$ & ****\\
			$\Xi_{c}^{'+}$  & $ 2578.2 \pm0.5 $ & $\frac{1}{2}^{+}$ & ***\\
			$\Xi_{c}^{'0}$  & $ 2578.7 \pm0.5 $ & $\frac{1}{2}^{+}$ & ***\\
			$\Xi_{c}(2645)^{+}$  &$ 2645.10 \pm0.30$ & $\frac{3}{2}^{+}$ & ***\\
			$\Xi_{c}(2645)^{0}$  &$2646.16 \pm0.25 $ & $\frac{3}{2}^{+}$ & ***\\
			$\Xi_{c}(2790)^{+}$  &$ 2791.9 \pm0.5$ & $\frac{1}{2}^{-}$ & ***\\
			$\Xi_{c}(2790)^{0}$  &$ 2793.9 \pm0.5$ & $\frac{1}{2}^{-}$ & ***\\
			$\Xi_{c}(2815)^{+}$  &$ 2816.51 \pm0.25$ & $\frac{3}{2}^{-}$ & ***\\
			$\Xi_{c}(2815)^{0}$  &$ 2819.79 \pm0.30$ & $\frac{3}{2}^{-}$ & ***\\
			$\Xi_{c}(2923)^{0}$  &$2923.04 \pm0.35 $ & $?^{?}$ & **\\
			$\Xi_{c}(2930)^{+}$  &$ 2942 \pm5$ & $?^{?}$ & **\\
			$\Xi_{c}(2930)^{0}$  &$ 2938.55 \pm0.30$ & $?^{?}$ & **\\
			$\Xi_{c}(2970)^{+}$  &$2964.3 \pm1.5 $ & $\frac{1}{2}^{+}$ & ***\\
			$\Xi_{c}(2970)^{0}$  &$2967.1 \pm1.7 $ & $\frac{1}{2}^{+}$ & ***\\
			$\Xi_{c}(3055)^{+}$  &$3055.9 \pm0.4 $ & $?^{?}$ & ***\\
			$\Xi_{c}(3080)^{+}$  &$3077.2 \pm0.4 $ & $?^{?}$ & ***\\
			$\Xi_{c}(3080)^{0}$  &$3079.9 \pm1.4 $ & $?^{?}$ & ***\\
			$\Xi_{c}(3123)^{+}$  &$3122.9 \pm1.3 $ & $?^{?}$ & *\\
			$\Omega_{c}^{0}$  &$ 2695.2 \pm1.7 $  & $\frac{1}{2}^{+}$ & ***\\
			$\Omega_{c}(2770)^{0}$  &$2765.9 \pm2.0 $ & $\frac{3}{2}^{+}$ & ***\\
			$\Omega_{c}(3000)^{0}$  &$3000.41 \pm0.22 $ & $?^{?}$ & ***\\
			$\Omega_{c}(3050)^{0}$  &$3050.19 \pm0.13$ & $?^{?}$ & ***\\
			$\Omega_{c}(3065)^{0}$  &$3065.54 \pm0.26 $ & $?^{?}$ & ***\\
			$\Omega_{c}(3090)^{0}$  &$3090.1 \pm0.5 $ & $?^{?}$ & ***\\
			$\Omega_{c}(3120)^{0}$  &$ 3119.1 \pm1.0$ & $?^{?}$ & ***\\
		\end{tabular}
	\end{ruledtabular}
\end{table}
Singly charmed baryons provide the best environment to investigate the dynamics of light quarks in the presence of a heavy charm quark. 
In recent years, a significant experimental effort has been made to measure the singly charmed baryons. Many experimental groups such as LHCb, Belle, BaBar, and CLEO, have provided data and are expected to produce more precise results in near future \cite{pdg2022}. 

The latest review of particle physics by Particle Data Group (PDG) \cite{pdg2022} confirms the six states of $\Lambda_c$ baryon with their respective spin and parity (see Table \ref{tab:table1}). 
But, $\Lambda_{c}(2765)^{+}$/$\Sigma_{c}(2765)^{+}$  is still a controversial state. It was first observed in  $ \Lambda_{c}\pi^{+}\pi^{-}$  decay channel by CLEO Collaboration \cite{artuso2001} and later confirmed by Belle  in $ \Sigma_{c}\pi $ decay mode \cite{abe2007}. We are uncertain of the identity of the observed state because both $ \Lambda_{c}^{+} $ and $ \Sigma_{c}^{+} $ can decay through these two channels.
However, Belle Collaboration \cite{abdesselam} predicts its isospin to be zero, and the particle is predicted as a state of $ \Lambda_{c} $.

For $ \Xi_{c} $ baryon family, states belonging to 1S and 1P wave with $J^{P}=\frac{1}{2}^{+}, \frac{1}{2}^{-},\frac{3}{2}^{-}$ have been well established. Besides of these states, six other states are also included in the PDG \cite{pdg2022} as shown in the Table \ref{tab:table1}. The spin and parity of these states, with the exception of  $ \Xi_{c}(2970) $  state, are still unknown. The $ \Xi_{c}(3055)^{+} $ and $ \Xi_{c}(3123)^{+} $ states were first observed by BaBar Collaboration in $ \Sigma_{c}(2455)^{++}K^{-} $ and $ \Sigma_{c}(2520)^{++}K^{-} $ channel respectively \cite{BABARCac_c}. Belle confirmed $ \Xi_{c}(3055)^{+} $ state, but no signal was found for $ \Xi_{c}(3123)^{+} $ state  \cite{BelleCAs_c}. The findings of the $ \Xi_{c}(3080) $  state was first reported by Belle \cite{Cas_c2970_1} and then verified by BaBar \cite{BABARCac_c}. In 2020, LHCb observed three excited $ \Xi_{c}^{0} $ resonances called  $ \Xi_{c}(2923)^{0} $, $ \Xi_{c}(2939)^{0} $, and $ \Xi_{c}(2965)^{0} $ in the $\Lambda_{c}^{+}K^{-}$ mass spectrum \cite{LHCb2020Cas_c}.  $ \Xi_{c}(2923)^{0} $ and  $ \Xi_{c}(2939)^{0} $ states are observed for the first time. This study indicates that the broad peak observed by Belle \cite{EPJCCas_c,EPJCCas_c1}  and BaBar  \cite{BABARCac_c1} for the $ \Xi_{c}(2930)^{0} $ state resolves into two separate peaks for the $ \Xi_{c}(2923)^{0} $ and $ \Xi_{c}(2939)^{0} $ states. But, $ \Xi_{c}(2965)^{0} $ state lies in the vicinity of previously observed state $ \Xi_{c}(2970)^{0} $  \cite{Cas_c2970_1,Cas_c2970_2,BABARCac_c}. Thus, 	further study is required to  establish whether or not the states $ \Xi_{c}(2965)^{0} $  and $ \Xi_{c}(2970)^{0} $  are equivalent. More recently in 2021, Belle reported first experimental determination of spin-parity of  $ \Xi_{c}(2970)^{+} $  using angular distribution of decay products in  chain $\Xi_{c}(2970)^{+}\rightarrow \Xi_{c}(2645)^{0}\pi^{+}\rightarrow \Xi_{c}^{+}\pi^{-}\pi^{+}$ and ratio of branching fraction of two decays $\Xi_{c}(2970)^{+}\rightarrow \Xi_{c}(2645)^{0}\pi^{+}/ \Xi_{c}'^{0}\pi^{+}$ \cite{Bellecas_c2970}. Their analysis favour $ J^{P}=\frac{1}{2}^{+} $ over other possibilities with the  zero spin of the light-quark degrees of freedom for $ \Xi_{c}(2970)^{+} $.

For $ \Sigma_{c}$, $ \Xi_{c}' $, and $ \Omega_{c} $ baryons, despite multiple theoretical and experimental attempts, only states belonging to 1S-wave with  $J^{P}=\frac{1}{2}^{+}$ and  $\frac{3}{2}^{+}$ have been discovered, and higher excited states still need to be established. So far, just one excited state of  $ \Sigma_{c}$ named $ \Sigma_{c}$(2800) has been discovered by Belle and BaBar collaborations in the channel of $\Lambda_{c}^{+}\pi$ \cite{Sigma_c2005,Sigma_c2008}. Its spin and parity are not identified yet. In 2017, LHCb declared the first observation of five narrow excited states of $ \Omega_{c}^{0} $ in $ \Xi_{c}^{+}K^{-} $channel:  $ \Omega_{c}(3000)^{0} $, $ \Omega_{c}(3050)^{0} $, $ \Omega_{c}(3065)^{0} $, $ \Omega_{c}(3090)^{0} $, and $ \Omega_{c}(3120)^{0} $  \cite{LHCbOmega_c}. Later, except for $ \Omega_{c}(3120)^{0} $, the other four states were confirmed by Belle collaboration \cite{BelleOmega_c}. \textbf{Recently, observation of two new excited states, $\Omega_{c}(3185)^{0}$  and $\Omega_{c}(3327)^{0}$  in the $\Xi_{c}^{+}K^{-}$ invariant-mass spectrum, was revealed by LHCb collaboration \cite{omegac3185and3327}.
}These latest findings motivate us to identify the spin and parity of these seven states of  $ \Omega_{c}^{0} $ baryon so that they can be fitted into their mass spectrum. To achieve this, the sufficient experimental information about the $\Lambda_c$ and $\Xi_c$ baryonic states can be used to study the nature of other singly charmed baryons, such as the $ \Sigma_{c}$, $ \Xi_{c}' $, and $ \Omega_{c} $ baryons.

The spectra of singly charmed baryons have been examined by numerous theoretical models, particularly quark model \cite{garcilazo2007,rbrt2008,valcarce2008,ebert2008, ebert2011,yoshida2015,shah2016,chen2017,wang2019,luo2023}, heavy hadron chiral perturbation theory  \cite{cheng2007,cheng2015,kawakami2019}, lattice QCD \cite{perezrubio2015}, light cone QCD sum rules \cite{hchen2017}, QCD sum rules \cite{1zhang2008, 2zhang2008, zwang2010,zwang2011, hchen2016, mao2017, zwang2018, zwang2021}, Regge phenomenology \cite{oudichhya2021} and relativistic flux tube  model \cite{ dwang2011, chen2015, PhysRevD.101.034016}. 
In Ref. \cite{Jenkins1996}, the authors studied the masses of baryons containing one heavy quark in a combined expansion in 1/$\mathbf{ m_{Q}} $ , 1/$\mathbf{ N_{c}} $ , and SU(3) flavour symmetry breaking.  

Five extremely narrow excited $ \Omega_{c} $  baryons that were recently detected were analyzed by the authors of  Ref. \cite{karliner2017}. As well as possible spin assignments and the relation between the masses of different states are examined.

In our previous work \cite{oudichhya2021}, we employed Regge phenomenology to calculate ground state and excited state masses of  $ \Omega_{c}^{0} $,  $ \Omega_{cc}^{+} $ and $ \Omega_{ccc}^{*++} $. However, this model is unable to predict all possible states of a system. We aim to uncover another model that is capable of predicting the entire spectrum. In ref. \cite{chen2015} authors have analytically derived linear Regge relation in a relativistic flux tube model for a heavy-light baryonic system. This relation is used to predict the complete spectrum of  $\Lambda_c$ and $\Xi_c$ baryons. But, they exclude the study of other singly charmed baryonic systems($\Sigma_c$, $\Xi'_c$, and $\Omega_c$), with vector diquark, due to the complexity of spin-dependent interactions.

In the present calculation, we apply linear Regge relation, introduced in ref. \cite{chen2015}, to all singly charmed baryons in the quark-diquark picture. Diquark, which is thought to be in its ground state, is assumed to excite orbitally or radially with respect to charm quark.  We incorporate spin-orbit interaction, spin-spin contact hyperfine interaction, and tensor interaction in the j-j coupling scheme to find spin-dependent splitting, and obtain the complete mass spectra of singly charmed baryons. We aim to identify the masses of fairly high orbital and radial excited states as well as assign possible quantum numbers to experimentally observed states of singly charmed baryons.

Following a brief overview of the experimental as well as theoretical progress, 
in Sec. II we discuss the details of the theoretical framework we have used to calculate the mass spectra of singly charmed baryons. It involves the derivation of the spin average mass formula in the relativistic flux tube model. Later, spin-dependent interactions are introduced. The parameters involved in this framework are calculated to obtain the mass spectra.
In Sec. III, the obtained results are discussed to assign the spin-parity to experimentally available states and to compare it with other theoretical predictions. 
Finally, we outline our conclusion in Section IV.

\section{Theoretical Framework}
The relativistic flux tube model \cite{PhysRevD.31.2910,lacourse1989,PhysRevD.45.4307,PhysRevD.48.417,PhysRevD.49.4675,Olsson1994,PhysRevD.51.3578,PhysRevD.53.4006,PhysRevD.60.074026} has achieved phenomenological success in explaining the  Regge trajectory behavior of hadrons. It is based on Nambu's idea of a dynamical gluonic string, which is responsible for the confinement of quarks within hadrons \cite{PhysRevD.10.4262}. This model has been extended in a variety of ways to study mesons \cite{PhysRevC.76.025206, PhysRevD.80.071502,Shan Hong-Yun_2010, PhysRevD.82.074003,jia2019regge, PhysRevD.101.014020}, baryons \cite{Chen Bing_2009,chen2015, PhysRevD.101.034016,song2022} as well as exotic hadrons \cite{PhysRevC.76.025206,PhysRevD.68.074007, PhysRevD.77.054020,PhysRevD.80.074021,Nandan:2016uce}. The model is also derived from QCD based Wilson area law\cite{brambilla1995}.

Singly charmed baryons are composed of charm quark($ c $) with two light quarks($ u,$  $ d $ or $  s $). According to heavy quark symmetry, the coupling between the $ c $ quark and the light quark is predicted to be weak \cite{isgur1991} and strong correlation between two light quarks ($ u,$  $ d $ or $  s $) permits the formation of a diquark. Apart from this, quark-diquark picture of baryons is supported by a number of theoretical arguments. A string model represents baryons as pieces of open strings connected at one common point\cite{Hooft}. This model predicts that a baryonic state in which three quarks are coupled to one another by three open strings that are joined at a single point in a Y-shape is unstable. One of the three arms will eventually vanish, releasing its energy into the excitation modes of the other two arms. The classically stable configuration is made up of one open string connecting two quarks at the end points and one quark travelling along the string. However, an attraction between two quarks into a $ \bar{3} $ bound state, one has a configuration of one quark at one end and a diquark at the other end of a single open string.  Moreover ref.\cite{Semay2008} concludes that the quark-diquark structure minimises baryon energy and is therefore preferred over the structure in which light quarks orbit around stationary heavy quark. The author in ref. \cite{Narodetskii2009} has shown that the singly heavy baryonic states with orbital angular momentum between a heavy quark and two light quarks are energetically favoured compared to the states with orbital angular momentum between two light quarks. This suggests that the states where two light quarks do not excite orbitally relative to each other and behave as a bound system, i.e., a diquark, are preferred. Inspired from these theoretical evidences, we take singly charmed baryons as a pair of diquark and charm quark.

 In the context of the relativistic flux tube(RFT) model, a gluonic field connecting diquark with charm quark is proposed to lie in a straight tube-like structure called a flux tube.
The color confinement is accomplished via this tube. The whole system of the charm quark, diquark, and flux tube, is constantly rotating with angular speed $ \omega $ around its center of mass. Along with the energy of a flux tube, this model also includes the angular momentum of a flux tube having string tension $ T $.

The relativistic Lagrangian $ \mathfrak{L} $, of the $ cqq  $ baryon in RFT model is  \cite{lacourse1989}
\begin{equation}
\label{eq:1}
\mathfrak{L}=\displaystyle\sum_{i=1} ^{2} \left[m_{i}(1-r_{i}^{2}\omega^{2})^{\frac{1}{2}}+T\int_{0}^{r_{i}}dr(1-r^{2}\omega^{2})^{\frac{1}{2}}\right],
\end{equation}
where  $m_1$ and $m_2$ can be accounted for current quark masses of  diquark and  charm quark, and $r_i$ denotes  its position from the centre of mass. For simplicity, we have chosen speed of light c=1, in natural units.

We write the total orbital angular momentum $ L $ as
\begin{equation}
	\label{eq:2}
	L=\frac{\partial \mathfrak{L}}{\partial \omega}=\displaystyle\sum_{i=1} ^{2} \left[\frac{m_{i}v_{i}^{2}}{\omega\sqrt{1-v_{i}^{2}}}+\frac{T}{\omega^{2}}\int_{0}^{v_{i}}\frac{v^{2}dv}{(1-v)^{2}}\right],
\end{equation}
where $ v_i =\omega r_i$  and  $ v=\omega r$.

The Hamiltonian $ H $, which is equivalent to the mass of the $ cqq $ baryon,  $\bar{M}$, is given by
\begin{equation}
	\label{eq:3}
		H=\bar{M}=\displaystyle\sum_{i=1} ^{2} \left[\frac{m_{i}}{\sqrt{1-v_{i}^{2}}}+\frac{T}{\omega}\int_{0}^{v_{i}}\frac{dv}{(1-v)^{2}}\right].
\end{equation}

We now define the effective mass of  diquark by $ m_d=m_1/\sqrt{1-v_{1}^{2}} $, and that of  charm quark by $ m_c=m_2/\sqrt{1-v_{2}^{2}} $, where $ v_{1} $ and $ v_{2} $ are speed of diquark and charm quark, respectively.
Then performing integration, Eq.(\ref{eq:2}) and  Eq.(\ref{eq:3}) simplifies to 
\begin{equation}
	\label{eq:4}
	L=\frac{1}{\omega}(m_{d}v_{1}^{2}+m_{c}v_{2}^{2})+\frac{T}{2\omega^{2}}\displaystyle\sum_{i=1} ^{2}(sin^{-1}v_{i}-v_{i}\sqrt{1-v_{i}^2}),
\end{equation}
and
\begin{equation}
	\label{eq:5}
\bar{M}= m_{d}+m_{c}+\frac{T}{\omega}\displaystyle\sum_{i=1} ^{2}sin^{-1}v_{i}.
\end{equation}

The boundary condition at the end of the flux tube with charm quark gives 
\begin{equation}
		\label{eq:6}
\frac{T}{\omega}=
\frac{m_c v_2}{\sqrt{1-v_{2}^{2}}}=m_c v_2[1+\frac{v_2^2}{2}+\frac{3v_2^4}{8}+...]\simeq m_c v_2,
\end{equation}
where higher order terms of $ v_{2} $ are neglected.

For singly charmed baryons $ m_{d}\ll m_{c}$. With this limit of a very small mass of diquark, we take the speed of light diquark $ v_{1}\approx 1 $ for approximation. Expanding  Eq. (\ref{eq:4}) and (\ref{eq:5}) in terms of $ v_{2} $ up to second order and using Eq. (\ref{eq:6}), mass relation can be obtained as  \cite{chen2015}

\begin{equation}
		\label{eq:7}
(\bar{M} -m_c)^2=\frac{\sigma}{2}L+(m_d+m_c v_{2}^{2}),
\end{equation}
where $\sigma=2\pi T$. This gives Regge-like relation between mass and angular momentum.
 \textbf{The method used in this study can be thought of as a semi-classical approximation of the quantized theory of strings as L is taken as the angular momentum quantum number. Despite the fact that the  we did not account for the quantum correction, the obtained Regge relation shows the basic features predicted by a fully quantum mechanical approach\cite{PhysRevD.49.4675, Olsson1994, PhysRevD.60.074026} such as linear nature of Regge trajectory and non zero intercept for L=0 state in $(L,(\bar{M}-m_c)^2)$  plane.}

Now, the distance between a heavy and light component of baryon can be given as 
\begin{equation}
\label{eq:8}
r=\frac{v_1+v_2}{\omega}=(v_1+v_2)\sqrt{\frac{8L}{\sigma}},
\end{equation}
where the relation between the angular speed of rotation of flux tube and orbital angular momentum, $ \omega=\sqrt{\sigma/8L} $, is obtained by combining Eq.(\ref{eq:4})  and (\ref{eq:7}).

\textbf{In Ref. \cite{PhysRevD.49.4675, Olsson1994, PhysRevD.60.074026}, the RFT model is solved for heavy-light mesons with a quantum mechanical approach, which gives a nearly straight leading Regge trajectory in $(L,(\bar{M}-m_c)^2)$  plane, which is followed by nearly parallel and equally spaced daughter trajectories. In our model, the singly charmed baryons are pictured as two body system with one heavy quark, the same as heavy-light mesons, where one quark is heavy. So, for singly charmed baryons with two-body picture, we will also get a nearly straight, parallel and equally spaced Regge trajectories.  In light of these quantum mechanical studies that show Regge trajectories in$(L,(\bar{M}-m_c)^2)$ plane with various values of $ n $ (radial excitation quantum number) to be parallel to one another, we extend our semi-classical relation (\ref{eq:7})and (\ref{eq:8}) for radially excited states by replacing $ L $ with ($ \lambda n+L $),  as}
	\begin{equation}
			\label{eq:10}
		(\bar{M}-m_c)^2=\frac{\sigma}{2}[\lambda n_{r}+L]+(m_d+m_c v_{2}^{2}),
	\end{equation}
and
\begin{equation}
		\label{eq:11}
	r=(v_1+v_2)\sqrt{\frac{8[\lambda n_{r}+L]}{\sigma}},
\end{equation}
\textbf{where $\lambda$ is unknown parameter that we extracted using experimental data.} Here, $ n_{r}=n-1 $ where $ n $ represents the principal quantum number of the baryon state.

The RFT model considers the quarks to be spinless; hence, we now incorporate spin-dependent interactions to get the complete mass spectra. The resulting mass takes the form
\begin{equation}
	\label{eq:13}
	M=\bar{M}+\Delta{M},
\end{equation} 
where spin average mass $ \bar{M} $ can be obtained from Eq.(\ref{eq:11}), and contribution to mass from spin-dependent interaction is given by
\begin{equation}
	\label{eq:14}
	\Delta{M}= H_{so}+ H_{t}+ H_{ss} .
\end{equation} 
Here, the first term is spin-orbit interaction, with the form
\begin{equation}
	\label{eq:15}
	\begin{split}
		H_{so}&=  [(\frac{2\alpha}{3r^3}-\frac{b'}{2r}) \frac{1}{m_d^2 }+\frac{4\alpha}{3r^3} \frac{1}{m_c m_d}]\mathbf{L}.\mathbf{S_d}\\
		& \hspace{0.3cm} +[(\frac{2\alpha}{3r^3}-\frac{b'}{2r}) \frac{1}{m_c^2 }+\frac{4\alpha}{3r^3} \frac{1}{m_c m_d}]\mathbf{L}.\mathbf{S_c}\\
		&=a_{1}\mathbf{L}.\mathbf{S_d}+a_{2}\mathbf{L}.\mathbf{S_c},
	\end{split}
\end{equation} 
which results from the short-range one-gluon exchange contribution and the long-range Thomas-precession term \cite{chen2022}. The second spin-dependent interaction
\begin{equation} 
	\label{eq:16}
	\begin{split}
		H_t&=\frac{4\alpha}{3r^3} \frac{1}{m_c m_d} [\frac{3(\mathbf{S_d}.\mathbf{r})(\mathbf{S_c}.\mathbf{r})}{r^2} -\mathbf{S_d}.\mathbf{S_c}]\\
		&=b\hat{B},
	\end{split}
\end{equation} 
arises from magnetic-dipole-magnetic-dipole color hyperfine interaction, is a tensor term. Here, $ \hat{B} $=$ 3(\mathbf{S_d}.\mathbf{r})(\mathbf{S_c}.\mathbf{r})/r^2 -\mathbf{S_d}.\mathbf{S_c} $.  The last term
\begin{equation}
	\label{eq:17}
	\begin{split}
		H_{ss}&=\frac{32\alpha\sigma_0^3}{9\sqrt{\pi} m_c m_d}e^{-\sigma_0^2 r^2 } \mathbf{S_d}.\mathbf{S_c}\\
		&=c\mathbf{S_d}.\mathbf{S_c},
	\end{split}
\end{equation}
is spin-spin contact hyperfine interaction. The parameters $ b'$ and $ \sigma_0 $  can be fixed using experimental data. $ \alpha $ is the coupling constant.  $ \mathbf{S_c} $ and $ \mathbf{S_d} $ denote spin of  charm quark and  diquark, respectively.\\
\begin{figure}[h]
 	\includegraphics[scale=0.33]{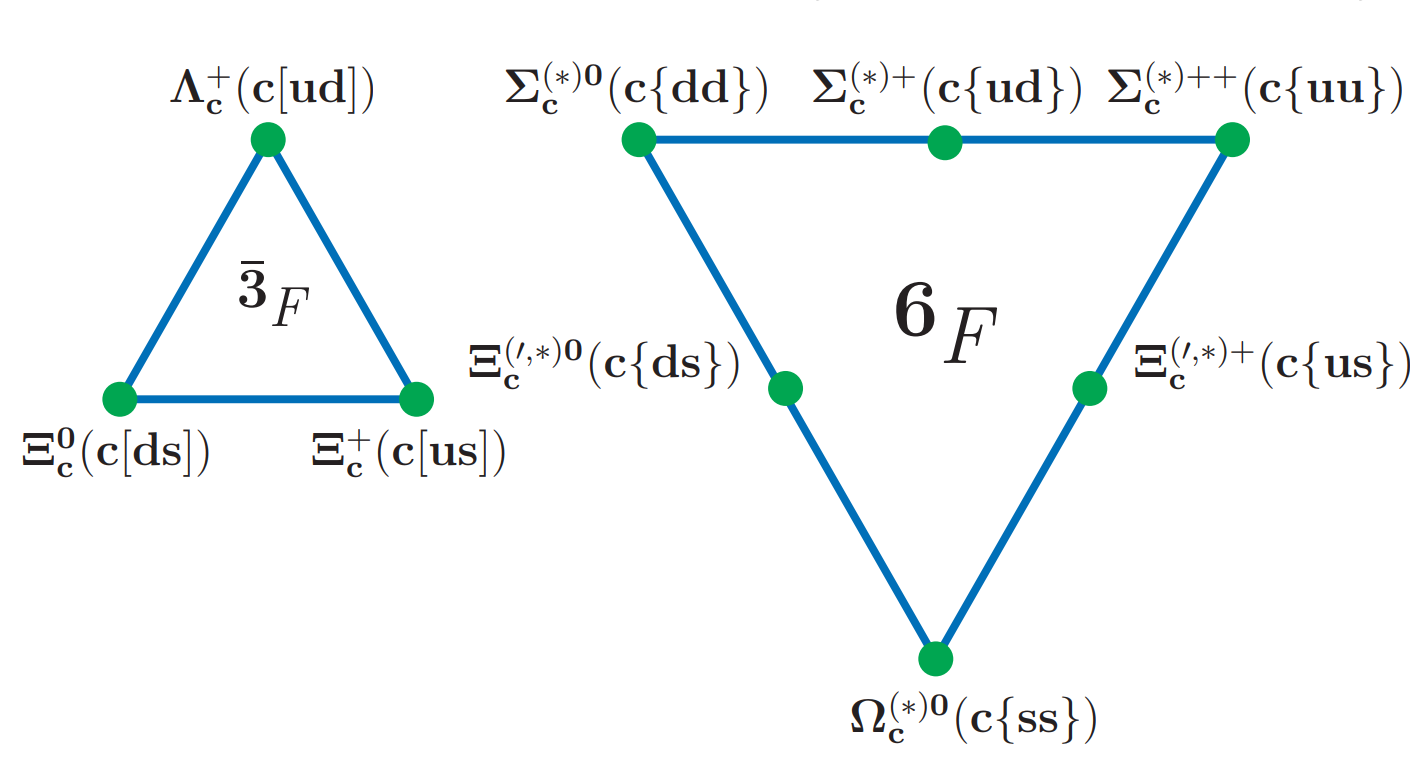}
 	\caption{\label{fig:1} Anti-triplet($ \bar{3}_{F} $) and sextet($ 6_{F} $) representation of singly charmed baryons.}
 \end{figure}
 As per Pauli's exclusion principle, the total symmetry of a diquark under an exchange of two quarks is antisymmetric. Diquark has a symmetric spatial state and an antisymmetric color state \cite{Selem:2006nd}. Flavor and spin states of diquark can either be symmetric or antisymmetric, such that $|flavor\rangle$$\times$$ |spin\rangle $ is symmetric. As shown in Fig.   {\ref{fig:1}}, flavor symmetries of light quarks organize singly charmed baryons into two groups, antisymmetric antitriplet ($ {\bar{3}}_{F} $) and symmetric sextet($ 6_{F} $),  as: $3\otimes3=6_{F}\oplus{\bar{3}}_{F} $.
The symmetric flavor state of a diquark requires a symmetric spin state, whereas the antisymmetric flavor state of a diquark requires an antisymmetric spin state. As a result, baryons that belong to the antitriplet($\Lambda_c$ and $\Xi_c$ baryons) have spin zero diquark (also known as scalar diquark $ [q_{1},q_{2}] $), whereas those that belong to the sextet($\Sigma_c$, $\Xi'_c$ and $\Omega_c$  baryons) have spin one diquark (also known as vector diquark $ \{q_{1},q_{2}\} $). Here, $ q_{1} $ and $  q_{2} $  are one of $ u,$  $ d $ or $  s $ quark.

There are two ways by which $ \mathbf{S_{c}} $, $ \mathbf{S_{d}} $ and $\mathbf{L}$ can couple to give total angular momentum $\mathbf{J}$. First is the L-S coupling scheme in which spin of diquark $ \mathbf{S_{d}} $ first couple with a spin of charm quark $ \mathbf{S_{c}} $ to form  $\mathbf{S}$ , later $\mathbf{S}$  couple with  $\mathbf{L}$ to give $\mathbf{J}$. The second one is the j-j coupling scheme, which is the dominant one in heavy quark limit,   where a spin of diquark $ \mathbf{S_{d}} $ first couple with $\mathbf{L}$ and results in total angular momentum of diquark $\mathbf{j}$, and then $\mathbf{j}$ couple with a spin of charm quark $ \mathbf{S_{c}} $ to give $\mathbf{J}$. Since $ m_{c}\gg m_{d}$ for single charmed baryons, we assume that heavy quark symmetry is followed. This allows us to refer to baryonic states as j-j coupling state $|J, j \rangle $, where both $\mathbf{j}$  and $\mathbf{J}$ are conserved.

For $\Lambda_c$ and $\Xi_c$ baryons, with scalar diquark($ S_{d}=0 $), both L-S and j-j coupling scheme gives identical states. Spin interactions are much simpler as only the second term of the spin-orbit interaction survives and tensor, as well as spin-spin contact hyperfine interactions, become zero. The expectation value of $ \mathbf{L}.\mathbf{S_c} $ in any coupling scheme for $ \Lambda_{c}  $and $\Sigma_{c}$  is given by 
\begin{equation}
	\langle\mathbf{L}.\mathbf{S_c}\rangle=\frac{1}{2}[J(J+1)-L(L+1)-S_c(S_c+1)].
\end{equation}

The spin interactions for $\Sigma_c$, $\Xi'_c$ and $\Omega_c$ baryons, with vector diquark($ S_{d}=1 $), are more complex than those for $\Lambda_c$ and $\Xi_c$ baryons. Detailed calculation of mass splitting operators, in j-j coupling scheme,  that are involved in spin-dependent interactions is shown in the appendix and results are listed in a Table \ref{tab:table7}.\\

\begin{table}
	\caption{\label{tab:table7}
	Matrix elements for spin interaction for different states for singly heavy baryon with vector diquark.
	}
	\begin{ruledtabular}
		
		\begin{tabular}{lrrrrccccccccc}
			
			$(L, J, j)$	&	$\langle \mathbf{S_{d}.L}\rangle$	&	$\langle \mathbf{S_{c}.L}\rangle$	&	$\langle\mathbf{\hat{B}}\rangle$	&	$\langle \mathbf{S_{d}.S_{c}}\rangle$	\\ 
			\hline
			$(S, 1/2, 1 )$ 	&	0	&	0	&	0	&	-1	\\ 
			$(S, 3/2, 1 )$ 	&	0	&	0	&	0	&	$ {1 }/{2}$	\\ 
			$(P, 1/2, 0 )$ 	&	-2	&	0	&	0	&	0	\\ 
			$(P, 1/2, 1 )$ 	&	-1	&	$-  {1 }/{2}$	&	-1	&	$-  {1 }/{2}$	\\ 
			$(P, 3/2, 1 )$ 	&	-1	&	$ {1 }/{4}$	&	$  {1 }/{2}$	&	$ {1 }/{4}$	\\ 
			$(P, 3/2, 2 )$ 	&	1	&	$-  {3 }/{4}$	&	$ {3 }/{10}$	&	$-  {3 }/{4}$	\\ 
			$(P, 5/2, 2 )$ 	&	1	&	$  {1 }/{2}$	&	$-  {1 }/{5}$	&	$  {1 }/{2}$	\\ 
			$(D, 1/2, 1 )$ 	&	-3	&	$-  {3 }/{2}$	&	-1	&	$  {1 }/{2}$	\\ 
			$(D, 3/2, 1 )$ 	&	-3	&	$  {3 }/{4}$	&	$  {1 }/{2}$	&	$- {1 }/{4}$	\\ 
			$(D, 3/2, 2 )$ 	&	-1	&	$-  {5 }/{4}$	&	$-  {1 }/{2}$	&	$-  {1 }/{4}$	\\ 
			$(D, 5/2, 2 )$ 	&	2	&	$-  {4 }/{3}$	&	$ {8 }/{21}$	&	$-  {2 }/{3}$	\\ 
			$(D, 5/2, 3 )$ 	&	-1	&	$  {5 }/{6}$	&	$  {1 }/{3}$	&	$  {1 }/{6}$	\\ 
			$(D, 7/2, 3 )$ 	&	2	&	1	&	$-  {2 }/{7}$	&	$  {1 }/{2}$	\\ 

		\end{tabular}
	\end{ruledtabular}
\end{table}  

\begin{table*}
	\caption{\label{tab:table2}
		Masses of $\Lambda_{c}^{+}$ baryon(in MeV).
	}
	\begin{ruledtabular}
		
		\begin{tabular}{clcccccccccc}
			$ (n, L, J, j) $& States	&	Present	&	PDG \cite{pdg2022}	&	 \cite{ebert2011}	&  \cite{chen2017}		&	  \cite{chen2015}	&	 \cite{shah2016}	&	 \cite{rbrt2008}	\\ 
			&\textit{$|nL, J^P\rangle$ }\\
			\hline
			
			(1, 0, 1/2, 0)   &  $|1S, 1/2^+\rangle$ 	&	2286.5	&	2286.46(0.14)	&	2286	&	2286	&	2286	&	2286	&	2268	\\ 
			(2, 0, 1/2, 0)   &  $|2S, 1/2^+\rangle$ 	&	2766.6	&	2766.6(0.24)	&	2769	&	2772	&	2766	&	2699	&	2791	\\ 
			(3, 0, 1/2, 0)   &  $|3S, 1/2^+\rangle$ 	&	3113.6	&		&	3130	&	3116	&	3112	&	3053	&		\\ 
			(4, 0, 1/2, 0)   &  $|4S, 1/2^+\rangle$ 	&	3399.9	&		&	3437	&		&	3397	&	3398	&		\\ 
			(1, 1, 1/2, 1)   &  $|1P, 1/2^-\rangle$ 	&	2592.3	&	2592.25(0.28)	&	2598	&	2614	&	2591	&	2629	&	2625	\\ 
			(1, 1, 3/2, 1)   &  $|1P, 3/2^-\rangle$ 	&	2628.1	&	2628.11(0.19)	&	2627	&	2639	&	2629	&	2612	&	2636	\\ 
			(2, 1, 1/2, 1)   &  $|2P, 1/2^-\rangle$ 	&	2989.6	&	$2939.6^{+1.3}_{-1.5}$	&	2983	&	2980	&	2989	&	2962	&	2816	\\ 
			(2, 1, 3/2, 1)   &  $|2P, 3/2^-\rangle$ 	&	3001.1	&		&	3005	&	3004	&	3000	&	2944	&	2830	\\ 
			(3, 1, 1/2, 1)   &  $|3P, 1/2^-\rangle$ 	&	3296.9	&		&	3303	&		&	3296	&	3295	&		\\ 
			(3, 1, 3/2, 1)   &  $|3P, 3/2^-\rangle$ 	&	3303.9	&		&	3322	&		&	3301	&	3276	&		\\ 
			(4, 1, 1/2, 1)   &  $|4P, 1/2^-\rangle$ 	&	3559.1	&		&	3588	&		&		&	3630	&		\\ 
			(4, 1, 3/2, 1)   &  $|4P, 3/2^-\rangle$ 	&	3564.3	&		&	3606	&		&		&	3610	&		\\ 
			(1, 2, 3/2, 2)   &  $|1D, 3/2^+\rangle$ 	&	2856.1	&	$2856.1^{+2.3}_{-6.0}$	&	2874	&	2843	&	2857	&	2873	&	2887	\\ 
			(1, 2, 5/2, 2)   &  $|1D, 5/2^+\rangle$ 	&	2881.6	&	2881.63(0.24)	&	2880	&	2851	&	2879	&	2849	&	2887	\\ 
			(2, 2, 3/2, 2)   &  $|2D, 3/2^+\rangle$ 	&	3189.6	&		&	3189	&		&	3188	&	3207	&	3073	\\ 
			(2, 2, 5/2, 2)   &  $|2D, 5/2^+\rangle$ 	&	3203.2	&		&	3209	&		&	3198	&	3179	&		\\ 
			(3, 2, 3/2, 2)   &  $|3D, 3/2^+\rangle$ 	&	3466.4	&		&	3480	&		&		&		&		\\ 
			(3, 2, 5/2, 2)   &  $|3D, 5/2^+\rangle$ 	&	3476.0	&		&	3500	&		&		&		&		\\ 
			(4, 2, 3/2, 2)   &  $|4D, 3/2^+\rangle$ 	&	3709.1	&		&	3747	&		&		&		&		\\ 
			(4, 2, 5/2, 2)   &  $|4D, 5/2^+\rangle$ 	&	3716.6	&		&	3767	&		&		&		&		\\ 
			(1, 3, 5/2, 3)   &  $|1F, 5/2^-\rangle$ 	&	3074.4	&		&	3097	&		&	3075	&	3116	&	2872	\\ 
			(1, 3, 7/2, 3)   &  $|1F, 7/2^-\rangle$ 	&	3097.2	&		&	3078	&		&	3092	&	3079	&		\\ 
			(2, 3, 5/2, 3)   &  $|2F, 5/2^-\rangle$ 	&	3369.0	&		&	3375	&		&		&		&		\\ 
			(2, 3, 7/2, 3)   &  $|2F, 7/2^-\rangle$ 	&	3384.0	&		&	3393	&		&		&		&		\\ 
			(3, 3, 5/2, 3)   &  $|3F, 5/2^-\rangle$ 	&	3622.9	&		&	3646	&		&		&		&		\\ 
			(3, 3, 7/2, 3)   &  $|3F, 7/2^-\rangle$ 	&	3634.3	&		&	3667	&		&		&		&		\\ 
			(1, 4, 7/2, 4)   &  $|1G, 7/2^+\rangle$ 	&	3265.9	&		&	3270	&		&	3267	&		&		\\ 
			(1, 4, 9/2, 4)   &  $|1G, 9/2^+\rangle$ 	&	3287.8	&		&	3284	&		&	3280	&		&		\\ 
			(2, 4, 7/2, 4)   &  $|2G, 7/2^+\rangle$ 	&	3533.1	&		&	3546	&		&		&		&		\\ 
			(2, 4, 9/2, 4)   &  $|2G, 9/2^+\rangle$ 	&	3549.1	&		&	3564	&		&		&		&		\\ 
			(1, 5, 9/2, 5)   &  $|1H, 9/2^-\rangle$ 	&	3438.9	&		&	3444	&		&		&		&		\\ 
			(1, 5,11/2, 5)   &  $|1H, 11/2^-\rangle$ 	&	3460.4	&		&	3460	&		&		&		&		\\ 

		\end{tabular}
	\end{ruledtabular}
\end{table*}  

\begin{table*}
	\caption{\label{tab:table3}
		Masses of $\Xi_{c}$ baryon(in MeV).}
	
	\begin{ruledtabular}
		
		\begin{tabular}{clcccccccccc}
			$ (n, L, J, j) $& States	&	Present	&	PDG \cite{pdg2022}	&	 \cite{ebert2011}	&	 \cite{chen2017}		&	  \cite{chen2015}		&		 \cite{rbrt2008}	\\ 
			&\textit{$|nL, J^P\rangle$ }\\
			\hline
			
			(1, 0, 1/2, 0)   &  $|1S, 1/2^+\rangle$ 	&	2470.4	&	2470.44(0.28)	&	2476	&	2470	&	2467	&	2466	\\ 
			(2, 0, 1/2, 0)   &  $|2S, 1/2^+\rangle$ 	&	2970.5	&	2967.10(1.7)	&	2959	&	2940	&	2959	&	2924	\\ 
			(3, 0, 1/2, 0)   &  $|3S, 1/2^+\rangle$ 	&	3342.8	&		&	3323	&	3265	&	3325	&		\\ 
			(4, 0, 1/2, 0)   &  $|4S, 1/2^+\rangle$ 	&	3653.1	&		&	3632	&		&	3629	&		\\ 
			(1, 1, 1/2, 1)   &  $|1P, 1/2^-\rangle$ 	&	2785.7	&	2793.90(0.5)	&	2792	&	2793	&	2779	&	2773	\\ 
			(1, 1, 3/2, 1)   &  $|1P, 3/2^-\rangle$ 	&	2823.9	&	2819.79(0.3)	&	2819	&	2820	&	2814	&	2783	\\ 
			(2, 1, 1/2, 1)   &  $|2P, 1/2^-\rangle$ 	&	3209.1	&		&	3179	&	3140	&	3195	&		\\ 
			(2, 1, 3/2, 1)   &  $|2P, 3/2^-\rangle$ 	&	3221.5	&		&	3201	&	3164	&	3204	&		\\ 
			(3, 1, 1/2, 1)   &  $|3P, 1/2^-\rangle$ 	&	3541.2	&		&	3500	&		&	3521	&		\\ 
			(3, 1, 3/2, 1)   &  $|3P, 3/2^-\rangle$ 	&	3548.9	&		&	3519	&		&	3525	&		\\ 
			(4, 1, 1/2, 1)   &  $|4P, 1/2^-\rangle$ 	&	3826.5	&		&	3785	&		&		&		\\ 
			(4, 1, 3/2, 1)   &  $|4P, 3/2^-\rangle$ 	&	3832.2	&		&	3804	&		&		&		\\ 
			
			(1, 2, 3/2, 2)   &  $|1D, 3/2^+\rangle$ 	&	3065.9	&	3055.9(0.4)	&	3059	&	3033	&	3055	&	3012	\\ 
			(1, 2, 5/2, 2)   &  $|1D, 5/2^+\rangle$ 	&	3093.3	&	3079.9(1.4)	&	3076	&	3040	&	3076	&	3004	\\ 
			(2, 2, 3/2, 2)   &  $|2D, 3/2^+\rangle$ 	&	3425.0	&		&	3388	&		&	3407	&		\\ 
			(2, 2, 5/2, 2)   &  $|2D, 5/2^+\rangle$ 	&	3439.7	&		&	3407	&		&	3416	&		\\ 
			(3, 2, 3/2, 2)   &  $|3D, 3/2^+\rangle$ 	&	3725.6	&		&	3678	&		&		&		\\ 
			(3, 2, 5/2, 2)   &  $|3D, 5/2^+\rangle$ 	&	3735.9	&		&	3699	&		&		&		\\ 
			(4, 2, 3/2, 2)   &  $|4D, 3/2^+\rangle$ 	&	3990.2	&		&	3945	&		&		&		\\ 
			(4, 2, 5/2, 2)   &  $|4D, 5/2^+\rangle$ 	&	3998.4	&		&	3965	&		&		&		\\ 
			
			(1, 3, 5/2, 3)   &  $|1F, 5/2^-\rangle$ 	&	3300.5	&		&	3278	&		&	3286	&		\\ 
			(1, 3, 7/2, 3)   &  $|1F, 7/2^-\rangle$ 	&	3325.0	&		&	3292	&		&	3302	&		\\ 
			(2, 3, 5/2, 3)   &  $|2F, 5/2^-\rangle$ 	&	3619.6	&		&	3575	&		&		&		\\ 
			(2, 3, 7/2, 3)   &  $|2F, 7/2^-\rangle$ 	&	3635.8	&		&	3592	&		&		&		\\ 
			(3, 3, 5/2, 3)   &  $|3F, 5/2^-\rangle$ 	&	3896.2	&		&	3845	&		&		&		\\ 
			(3, 3, 7/2, 3)   &  $|3F, 7/2^-\rangle$ 	&	3908.6	&		&	3865	&		&		&		\\ 
			
			(1, 4, 7/2, 4)   &  $|1G, 7/2^+\rangle$ 	&	3507.7	&		&	3469	&		&	3490	&		\\ 
			(1, 4, 9/2, 4)   &  $|1G, 9/2^+\rangle$ 	&	3531.3	&		&	3483	&		&	3503	&		\\ 
			(2, 4, 7/2, 4)   &  $|2G, 7/2^+\rangle$ 	&	3798.2	&		&	3745	&		&		&		\\ 
			(2, 4, 9/2, 4)   &  $|2G, 9/2^+\rangle$ 	&	3815.5	&		&	3763	&		&		&		\\ 
			
			(1, 5, 9/2, 5)   &  $|1H, 9/2^-\rangle$ 	&	3695.6	&		&	3643	&		&		&		\\ 
			(1, 5,11/2, 5)   &  $|1H, 11/2^-\rangle$ 	&	3719.0	&		&	3658	&		&		&		\\ 
%
%
%
%
		\end{tabular}
	\end{ruledtabular}
\end{table*}  

\subsection{Mass spectra of  $\Lambda_c$ and $\Xi_c$ baryons}

Experimental research has revealed a wide variety of $\Lambda_c$  baryonic states. This gives us context for fixing unidentified parameters in our model. The parameters $ m_c $, $ v_1 $, $ v_2 $, $\lambda$, $\alpha$, and $b'$, can be fixed from experimentally available states of $\Lambda_c$, are taken as common parameters for all singly charmed baryons to ensure consistency in the model. The remaining parameters, such as the diquark mass  $ m_d $, string tension $\sigma$, and $\sigma_{0}$, depends on the system under consideration.  The ultra-relativistic nature of the light diquark leads us to believe that $ v_1 $ $\approx$ 1. The spin-averaged mass of the $ nL $-wave is \cite{rai2008} 
\begin{equation}
	\bar{M}_{nL}=\frac{\sum_{J}(2J+1)M_{J}}{\sum_{J}(2J+1)},
\end{equation}
where the summation is taken over all possible states of $ nL $-wave with spin $ J $ and mass $ M_{J} $. The spin average mass of the $ 1S $, $ 1P $, and $ 1D  $ states is calculated using the corresponding experimental states of $\Lambda_c^{+}$ baryons, which is then utilised to determine the parameters $ m_c $= 1.448 GeV, $\sigma_{\Lambda_{c}}$ = 1.323  GeV$^2 $, and $ m_{d_{[u,d]}}$  $+m_c v_2^2=0.838$ GeV. We fix the velocity charm quark equal to 0.48 by comparing the mass of charm quark   $ m_c $ to its current-quark mass 1.27 ± 0.025 GeV  \cite{pdg2022}. This yields $ m_{d_{[u,d]}}$ = 0.503 GeV. The experimental mass of the $\Lambda_c(2765)^+$ state and the predicted mass of the $|2S, 1/2^+\rangle$  state from the relativistic quark model in ref. \cite{ebert2011} are highly comparable. So it makes sense to accept this conclusion and adapt it to  extract $\lambda=1.565$.  For $\Lambda_c$ baryon,  $ H_t $ and $ H_{ss} $ becomes zero. So, we fix $\alpha$=0.426 and $b'$=-0.076 GeV$ ^2 $ in eq.(\ref{eq:15}) using splitting in $ 1P  $ and $ 1D  $ wave. 
Utilizing above parameters mass spectra of $\Lambda_c$ baryon with comparison with other Refs. \cite{ebert2011,chen2017,chen2015,shah2016,rbrt2008}  is shown  in Table \ref{tab:table1}.
For the $\Xi_{c}$  baryons,  with the spin-averaged masses of $ 1S $ and $ 1P $, which we calculate using experimentally detected states, we extract the mass of diquark $ m_{d_{[d,s]}}$=0.687 GeV and $\sigma_{\Xi_{c}}$=1.625 GeV$^2$. With these parameters, we determine the masses of the $\Xi_{c}$ baryonic states, which are given in Table \ref{tab:table2}. \textbf{Our predicted mass for first radially excited state $|2S, 1/2^+\rangle$ of $\Xi_{c}$ baryon is 2970.5 MeV, which is only 3.4 MeV different from the experimentally detected mass of 2967.10 MeV.   This confirms that the extracted value of $\lambda=1.565$ from the experimental data of  $\Lambda_c$ baryon is reliable.}

\subsection{Mass spectra of $\Sigma_c$, $\Xi'_c$ and $\Omega_c$ baryons}

Using an experimental spin average mass of $ 1S $ wave of these three systems, we first find a mass of diquark  $m_{d_{\{u,u\}}}=0.714$ GeV, $m_{d_{\{d,s\}}}=0.841$ GeV and $m_{d_{\{s,s\}}}=0.959$ GeV. In spin-dependent interactions for S-wave, only spin-spin contact hyperfine interaction contributes. Thus, the parameter involved in this interaction,  $ \sigma_{0} $, is calculated using splitting in $ 1S $ wave in respective system as $\sigma_{0_{\Sigma_c}}=0.373$ GeV, $\sigma_{0_{\Xi'_c}}=0.400$ GeV and $\sigma_{0_{\Omega_c}}=0.425$ GeV. We take a spin average mass of $ 2S  $ wave of  $\Sigma_c$ as input from Ref. \cite{ebert2011} to fix $\lambda$ = 1.299 for these systems.
\begin{table}[!h]
	\caption{\label{tab:table8}
		In relativistic quark model, the mass of diquark and slope of Regge trajectory in (L, $ (\bar{M}-m_{c})^{2} $) plane \cite{ebert2011,chen2018}. This data shows that within a singly charmed baryonic family, the slope of the Regge trajectory in the (L, $ (\bar{M}-m_{c})^{2} $) plane increases along with the mass of diquark .}
	\begin{ruledtabular}
		
		\begin{tabular}{ccc}
			
			Baryon	&	$ m_{d} $(GeV)	&	Regge slope(GeV$ ^{-2} $)\\ 
			\hline
			
			$\Lambda_{c}$& 0.710 &	0.615\\
			$\Sigma_c$& 0.909 & 0.683\\
			$ \Xi_{c} $& 0.948	& 0.711\\
			$ \Xi_{c}' $& 1.069 & 0.752\\
			$ \Omega_{c} $& 1.203 & 0.812\\
		\end{tabular}
	\end{ruledtabular}
\end{table}  

\begin{table*}
	\caption{\label{tab:table4}
		Masses of $\Sigma_c$ baryon(in MeV).}
	
	\begin{ruledtabular}
		
		\begin{tabular}{clcccccccccc}
			$ (n, L, J, j) $& States	&	Present	&	PDG \cite{pdg2022}	&	 \cite{ebert2011}	&	  \cite{chen2017}		&	 \cite{shah2016}	&		 \cite{rbrt2008}	\\ 
			&\textit{$|nL, J^P\rangle$ }\\
			\hline
			
			(1, 0, 1/2, 1)   &  $|1S, 1/2^+\rangle$ 	&	2454.0	&	2453.97(0.14)	&	2443	&	2456	&	2454	&	2455	\\ 
			(1, 0, 3/2, 1)   &  $|1S, 3/2^+\rangle$ 	&	2518.4	&	$2518.41^{+0.22}_{-0.18}$	&	2519	&	2515	&	2530	&	2519	\\ 
			(2, 0, 1/2, 1)   &  $|2S, 1/2^+\rangle$ 	&	2917.9	&		&	2901	&	2850	&	3016	&	2958	\\ 
			(2, 0, 3/2, 1)   &  $|2S, 3/2^+\rangle$ 	&	2927.6	&		&	2936	&	2876	&	3069	&	2995	\\ 
			(3, 0, 1/2, 1)   &  $|3S, 1/2^+\rangle$ 	&	3252.3	&		&	3271	&	3091	&	3492	&		\\ 
			(3, 0, 3/2, 1)   &  $|3S, 3/2^+\rangle$ 	&	3253.7	&		&	3293	&	3109	&	3525	&		\\ 
			(1, 1, 1/2, 0)   &  $|1P,1/2^-\rangle$ 	&	2727.8	&		&	2713	&	2702	&	2890	&	2748	\\ 
			(1, 1, 1/2, 1)   &  $|1P, 1/2^-\rangle$ 	&	2749.3	&		&	2799	&	2765	&	2906	&	2768	\\ 
			(1, 1, 3/2, 1)   &  $|1P, 3/2^-\rangle $ 	&	2800.1	&	$2801^{+4}_{-6}$	&	2773	&	2785	&	2860	&	2763	\\ 
			(1, 1, 3/2, 2)   &  $|1P, 3/2^-\rangle $ 	&	2872.1	&		&	2798	&	2798	&	2875	&	2776	\\ 
			(1, 1, 5/2, 2)   &  $|1P, 5/2^-\rangle$ 	&	2908.5	&		&	2789	&	2790	&	2835	&	2790	\\ 
			(2, 1, 1/2, 0)   &  $|2P,1/2^-\rangle$ 	&	3134.8	&		&	3125	&	2971	&	3352	&		\\ 
			(2, 1, 1/2, 1)   &  $|2P, 1/2^-\rangle$ 	&	3149.2	&		&	3172	&	3018	&	3369	&		\\ 
			(2, 1, 3/2, 1)   &  $|2P, 3/2^-\rangle $ 	&	3164.3	&		&	3151	&	3036	&	3318	&		\\ 
			(2, 1, 3/2, 2)   &  $|2P, 3/2^-\rangle $ 	&	3201.5	&		&	3172	&	3044	&	3335	&		\\ 
			(2, 1, 5/2, 2)   &  $|2P, 5/2^-\rangle$ 	&	3212.8	&		&	3161	&	3040	&	3290	&		\\ 

		\end{tabular}
	\end{ruledtabular}
\end{table*}  

\begin{table*}
	\caption{\label{tab:table5}
		Masses of $\Xi'_c$ baryon(in MeV).}
	
	\begin{ruledtabular}
		
		\begin{tabular}{clcccccccccc}
			$ (n, L, J, j) $& States	&	Present	&	PDG \cite{pdg2022}	&	 \cite{ebert2011}	&	  \cite{chen2017}		&		 \cite{rbrt2008}	\\ 
			&\textit{$|nL, J^P\rangle$ }\\
			\hline

			(1, 0, 1/2, 1) &  $|1S, 1/2^+\rangle$ 	&	2578.7	&	2578.70(0.5)	&	2579	&	2579	&	2594	\\ 
			(1, 0, 3/2, 1) &  $|1S, 3/2^+\rangle$ 	&	2646.2	&	2646.16(0.25)	&	2649	&	2649	&	2649	\\ 
			(2, 0, 1/2, 1) &  $|2S, 1/2^+\rangle$ 	&	3049.3	&	3055.9(0.4)	&	2983	&	2977	&		\\ 
			(2, 0, 3/2, 1) &  $|2S, 3/2^+\rangle$ 	&	3058.9	&	3055.9(0.4)	&	3026	&	3007	&		\\ 
			(3, 0, 1/2, 1) &  $|3S, 1/2^+\rangle$ 	&	3393.1	&		&	3377	&	3215	&		\\ 
			(3, 0, 3/2, 1) &  $|3S, 3/2^+\rangle$ 	&	3394.5	&		&	3396	&	3236	&		\\ 
			
			(1, 1, 1/2, 0) &  $|1P,1/2^-\rangle$ 	&	2873.3	&		&	2854	&	2839	&	2855	\\ 
			(1, 1, 1/2, 1) &  $|1P, 1/2^-\rangle$ 	&	2886.4	&	2923.04(0.35)	&	2936	&	2900	&		\\ 
			(1, 1, 3/2, 1) &  $|1P, 3/2^-\rangle $ 	&	2937.9	&	2938(0.3)	&	2912	&	2921	&	2866	\\ 
			(1, 1, 3/2, 2) &  $|1P, 3/2^-\rangle $ 	&	2992.9	&		&	2935	&	2932	&		\\ 
			(1, 1, 5/2, 2) &  $|1P, 5/2^-\rangle$ 	&	3030.5	&		&	2929	&	2927	&	2989	\\ 
			
			(2, 1, 1/2, 0) &  $|2P,1/2^-\rangle$ 	&	3282.0	&		&	3267	&	3094	&		\\ 
			(2, 1, 1/2, 1) &  $|2P, 1/2^-\rangle$ 	&	3291.9	&		&	3313	&	3144	&		\\ 
			(2, 1, 3/2, 1) &  $|2P, 3/2^-\rangle $ 	&	3307.3	&		&	3293	&	3172	&		\\ 
			(2, 1, 3/2, 2) &  $|2P, 3/2^-\rangle $ 	&	3335.4	&		&	3311	&	3165	&		\\ 
			(2, 1, 5/2, 2) &  $|2P, 5/2^-\rangle$ 	&	3347.2	&		&	3303	&	3170	&		\\ 
			
			
			
			
			(1, 2, 1/2, 1) &  $|1D, 1/2^+\rangle$ 	&	3157.1	&	3122.9	&	3163	&	3075	&		\\ 
			(1, 2, 3/2, 1) &  $|1D, 3/2^+\rangle$ 	&	3189.8	&		&	3160	&	3089	&		\\ 
			(1, 2, 3/2, 2) &  $|1D, 3/2^+\rangle$ 	&	3207.1	&		&	3167	&	3081	&		\\ 
			(1, 2, 5/2, 3) &  $|1D, 5/2^+\rangle$ 	&	3236.6	&		&	3153	&	3091	&		\\ 
			(1, 2, 5/2, 2) &  $|1D, 5/2^+\rangle $ 	&	3279.0	&		&	3166	&	3077	&		\\ 
			(1, 2, 7/2, 3) &  $|1D, 7/2^+\rangle$ 	&	3304.6	&		&	3147	&	3078	&		\\

		\end{tabular}
	\end{ruledtabular}
\end{table*}  

\begin{table*}
	\caption{\label{tab:table6}
		Masses of $\Omega_c$ baryon(in MeV).}
	
	\begin{ruledtabular}
		
		\begin{tabular}{clcccccccccc}
			$ (n, L, J, j) $ &	States	&	Present	&	PDG \cite{pdg2022}	&	 \cite{ebert2011}	&	 \cite{oudichhya2021}		&		 \cite{rbrt2008}	&	 \cite{shah2016}	\\ 
			&\textit{$|nL, J^P\rangle$ }\\
			\hline
			
			(1, 0, 1/2, 1) & $|1S, 1/2^+\rangle$ 	&	2695.2	&	2695.2(1.7)	&	2698	&	2702	&	2718	&	2695	\\ 
			(1, 0, 3/2, 1) & $|1S, 3/2^+\rangle$ 	&	2765.9	&	2765.9(2.0)	&	2765	&	2772	&	2776	&	2745	\\ 
			(2, 0, 1/2, 1) & $|2S, 1/2^+\rangle$ 	&	3171.2	&		&	3088	&	3164	&	3152	&	3164	\\ 
			(2, 0, 3/2, 1) & $|2S, 3/2^+\rangle$ 	&	3180.5	&\textbf{3185.1(1.7)\cite{omegac3185and3327}}		&	3123	&	3197	&	3190	&	3197	\\ 
			(3, 0, 1/2, 1) & $|3S, 1/2^+\rangle$ 	&	3522.4	&		&	3489	&	3566	&		&	3561	\\ 
			(3, 0, 3/2, 1) & $|3S, 3/2^+\rangle$ 	&	3523.6	&		&	3510	&	3571	&		&	3580	\\ 
			
			(1, 1, 1/2, 0) & $|1P,1/2^-\rangle$ 	&	3003.2	&	3000.41(0.22)	&	2966	&		&	2977	&	3041	\\ 
			(1, 1, 1/2, 1) & $|1P, 1/2^-\rangle$ 	&	3010.7	&	3050.19(0.13)	&	3055	&		&	2990	&	3050	\\ 
			(1, 1, 3/2, 1) & $|1P, 3/2^-\rangle $ 	&	3062.8	&	3065.54(0.26)	&	3029	&	3049	&	2986	&	3024	\\ 
			(1, 1, 3/2, 2) & $|1P, 3/2^-\rangle $ 	&	3106.6	&	3090.10(0.5)	&	3054	&		&	2994	&	3033	\\ 
			(1, 1, 5/2, 2) & $|1P, 5/2^-\rangle$ 	&	3145.3	&	3119.10(1.0)	&	3051	&	3055	&	3014	&	3010	\\ 
			
			(2, 1, 1/2, 0) & $|2P,1/2^-\rangle$ 	&	3414.3	&		&	3384	&		&		&	3427	\\ 
			(2, 1, 1/2, 1) & $|2P, 1/2^-\rangle$ 	&	3421.3	&		&	3435	&		&		&	3436	\\ 
			(2, 1, 3/2, 1) & $|2P, 3/2^-\rangle $ 	&	3436.8	&		&	3415	&	3408	&		&	3408	\\ 
			(2, 1, 3/2, 2) & $|2P, 3/2^-\rangle $ 	&	3459.1	&		&	3433	&		&		&	3417	\\ 
			(2, 1, 5/2, 2) & $|2P, 5/2^-\rangle$ 	&	3471.1	&		&	3427	&	3393	&		&	3393	\\ 
			
			(3, 1, 1/2, 0) & $|3P,1/2^-\rangle$ 	&	3731.3	&		&	3717	&		&		&	3813	\\ 
			(3, 1, 1/2, 1) & $|3P, 1/2^-\rangle$ 	&	3737.2	&		&	3754	&		&		&	3823	\\ 
			(3, 1, 3/2, 1) & $|3P, 3/2^-\rangle $ 	&	3745.7	&		&	3737	&	3732	&		&	3793	\\ 
			(3, 1, 3/2, 2) & $|3P, 3/2^-\rangle $ 	&	3761.7	&		&	3752	&		&		&	3803	\\ 
			(3, 1, 5/2, 2) & $|3P, 5/2^-\rangle$ 	&	3768.7	&		&	3744	&	3700	&		&	3777	\\ 
			
			
			%
			(1, 2, 1/2, 1) & $|1D, 1/2^+\rangle$ 	&	3289.7	&		&	3287	&		&		&	3354	\\ 
			(1, 2, 3/2, 1) & $|1D, 3/2^+\rangle$ 	&	3323.5	&\textbf{3327.1(1.2)\cite{omegac3185and3327}}		&	3282	&		&		&	3325	\\ 
			(1, 2, 3/2, 2) & $|1D, 3/2^+\rangle$ 	&	3333.8	&	\textbf{3327.1(1.2)\cite{omegac3185and3327}}	&	3298	&		&		&	3335	\\ 
			(1, 2, 5/2, 3) & $|1D, 5/2^+\rangle$ 	&	3364.1	&		&	3286	&	3360	&	3196	&	3299	\\ 
			(1, 2, 5/2, 2) & $|1D, 5/2^+\rangle $ 	&	3396.5	&		&	3297	&		&		&	3308	\\ 
			(1, 2, 7/2, 3) & $|1D, 7/2^+\rangle$ 	&	3422.9	&		&	3283	&	3314	&		&	3276	\\ 

		\end{tabular}
	\end{ruledtabular}
\end{table*}  

\begin{figure}
	\hspace{-2.5cm}
	\includegraphics[scale=0.28]{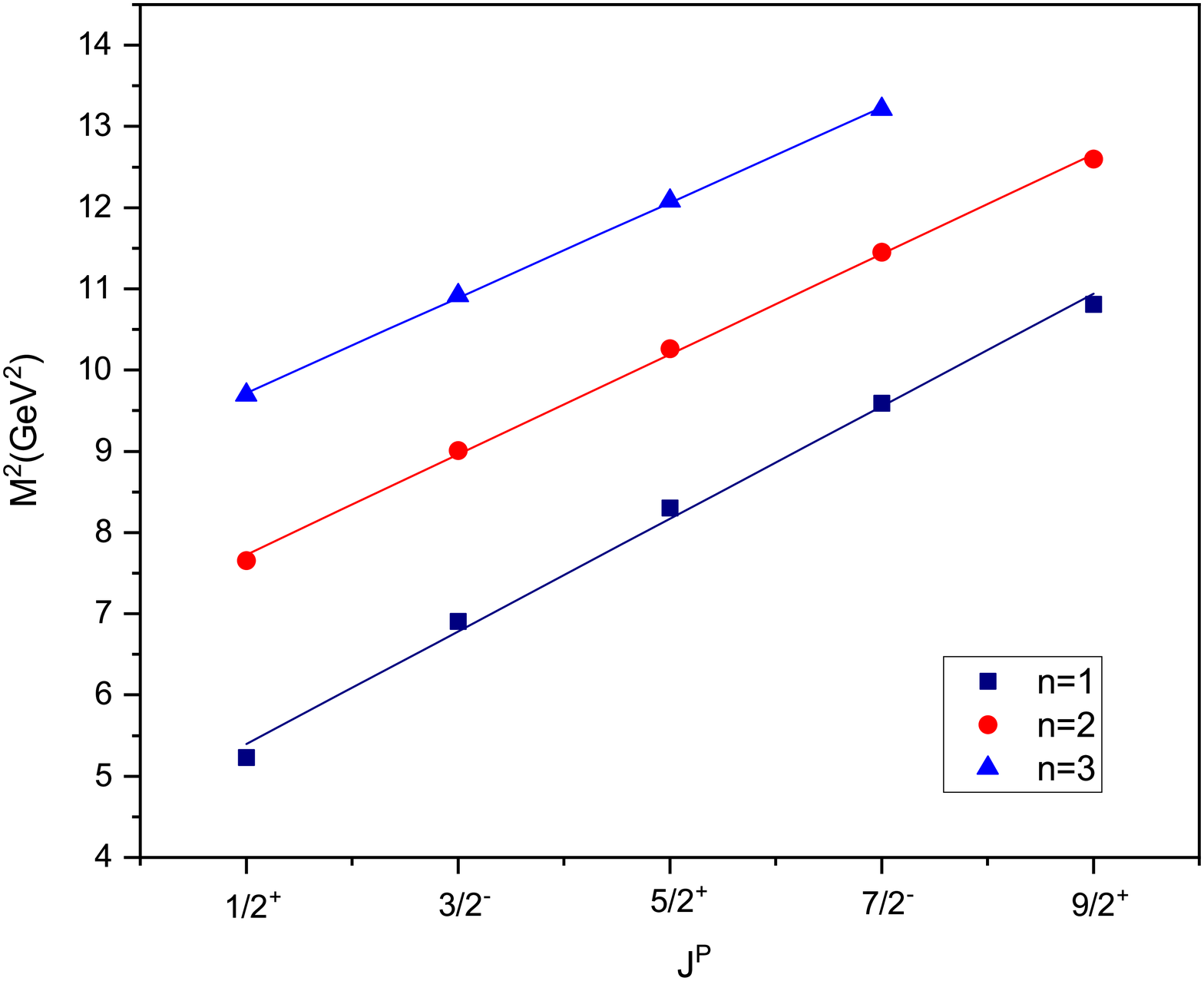}
	\caption{\label{fig:2}{Regge trajectory in the ($J,M^{2}$) plane for $\Lambda_{c}$ baryon for natural parity states}}
\end{figure}

\begin{figure}
	\hspace{-2.5cm}
	\includegraphics[scale=0.28]{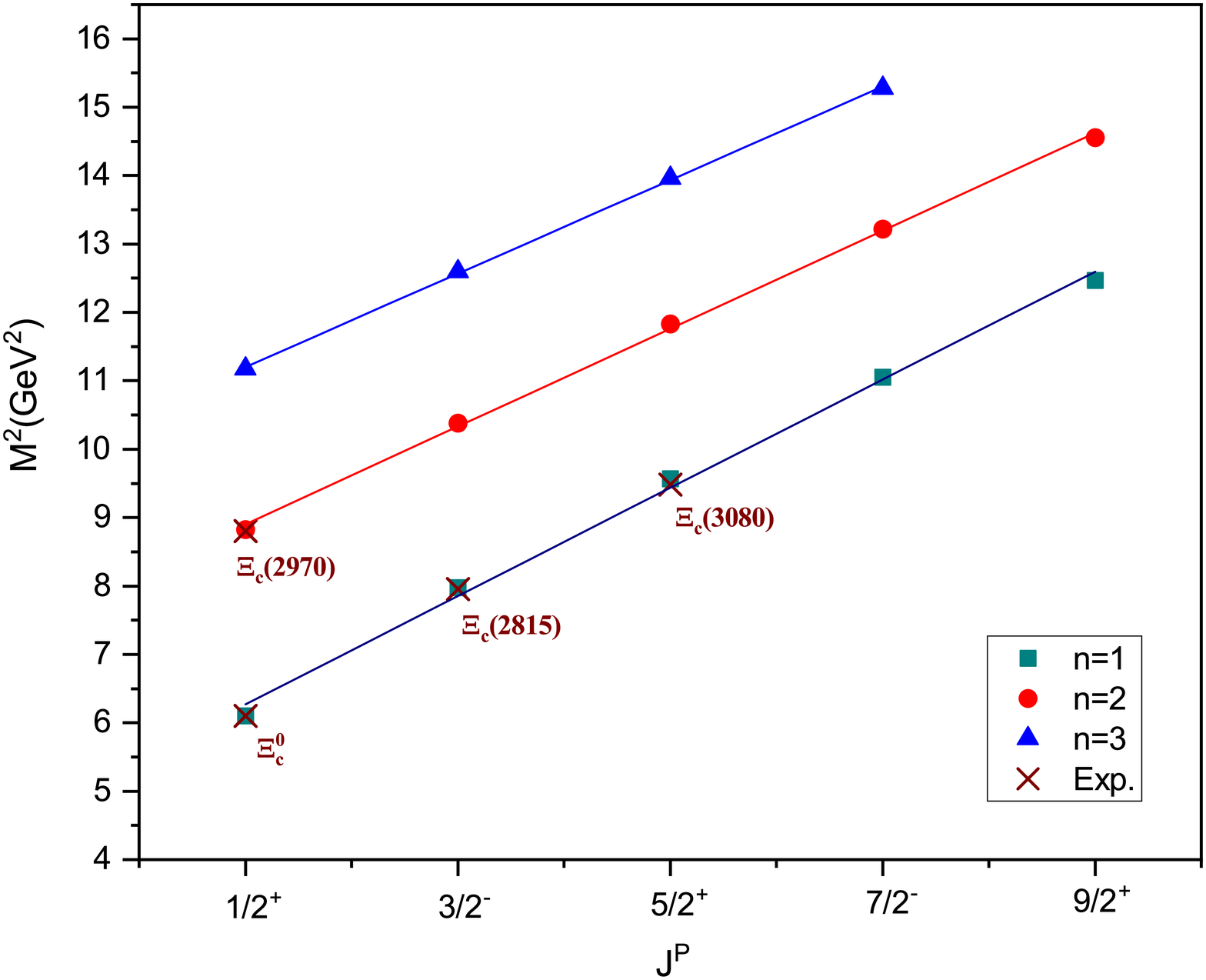}
	\caption{\label{fig:3}{Regge trajectory in the ($J,M^{2}$) plane for $\Xi_{c}$ baryon for natural parity states}}
\end{figure}

\begin{figure}
	\hspace{-2.5cm}
	\includegraphics[scale=0.28]{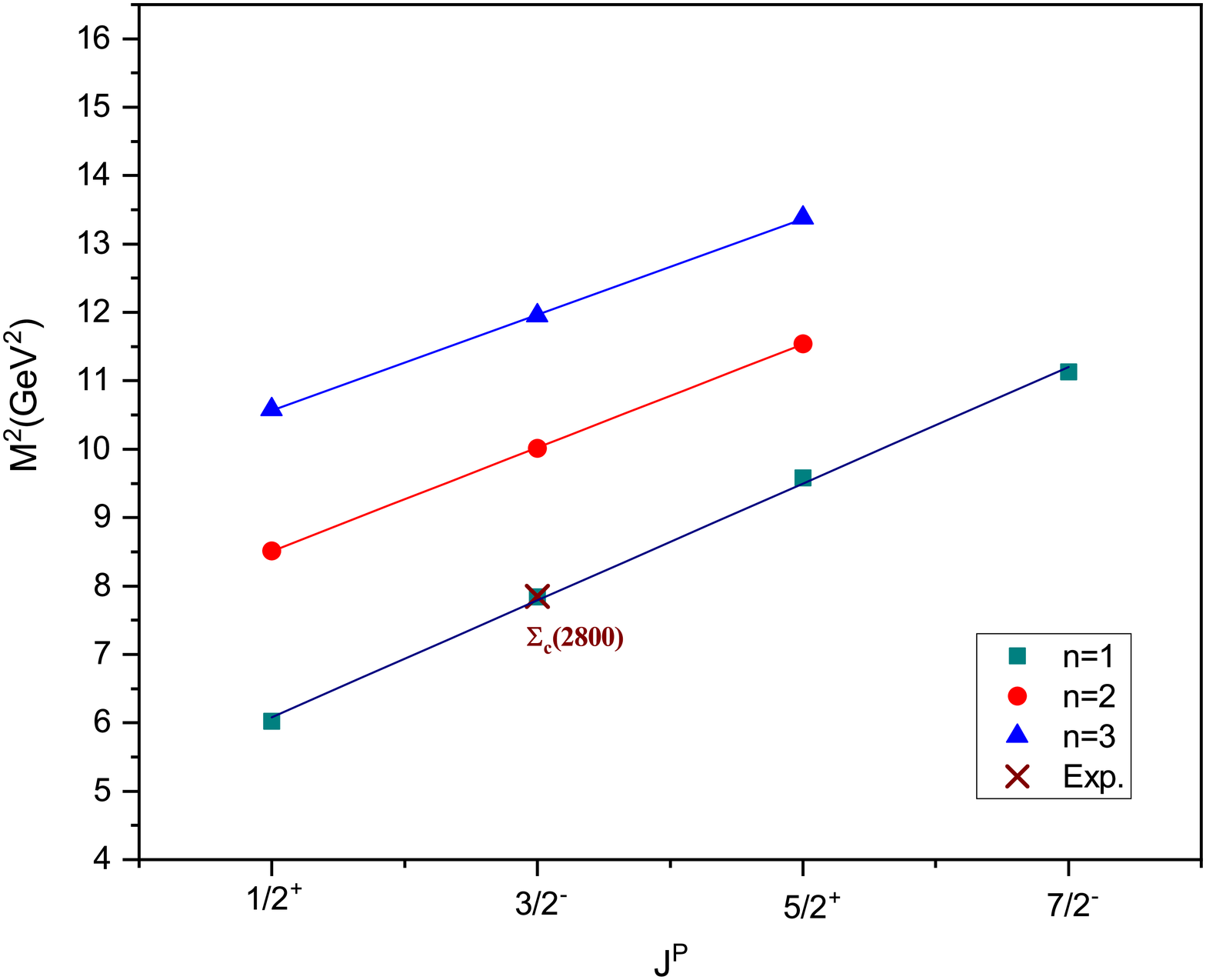}
	\caption{\label{fig:4}{Regge trajectory in the ($J,M^{2}$) plane for $\Sigma_{c}$ baryon for natural parity states}}
\end{figure}

\begin{figure}
	\hspace{-2.5cm}
	\includegraphics[scale=0.28]{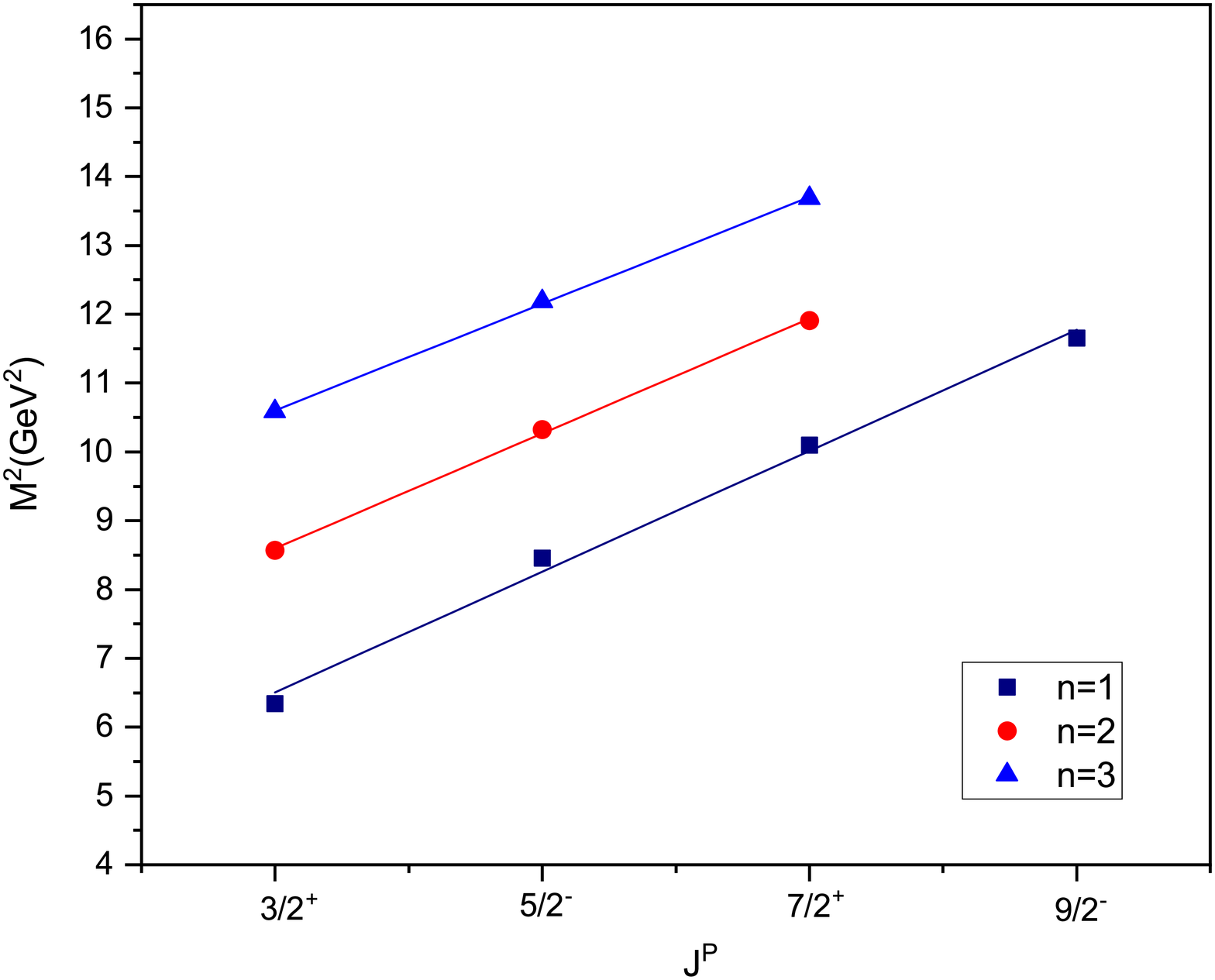}
	\caption{\label{fig:5}{Regge trajectory in the ($J,M^{2}$) plane for $\Sigma_{c}$ baryon for unnatural parity states}}
\end{figure}

\begin{figure}
	\hspace{-2.5cm}
	\includegraphics[scale=0.28]{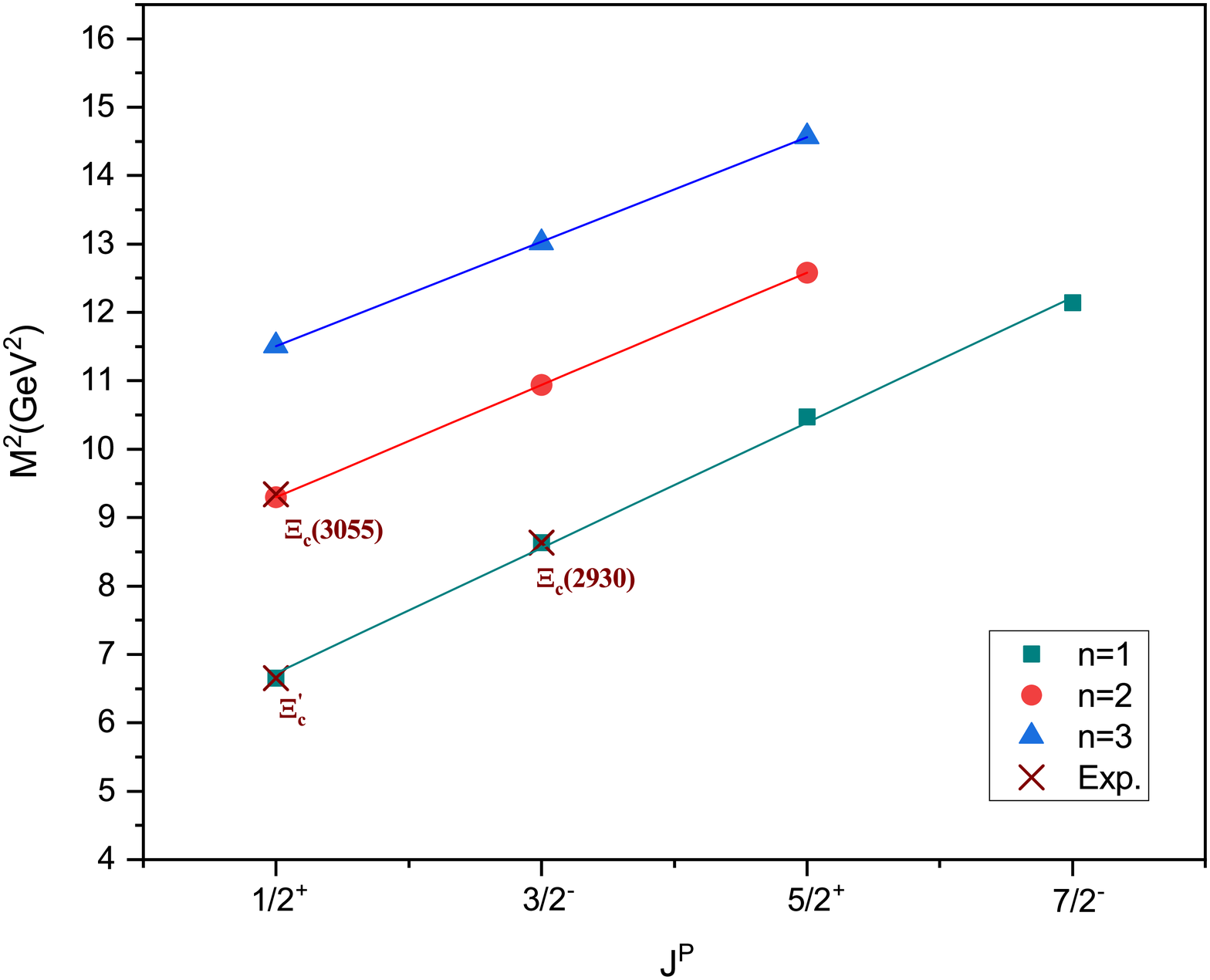}
	\caption{\label{fig:6}{Regge trajectory in the ($J,M^{2}$) plane for $\Xi_{c}^{'}$ baryon for natural parity states}}
\end{figure}

\begin{figure}
	\hspace{-2.5cm}
	\includegraphics[scale=0.28]{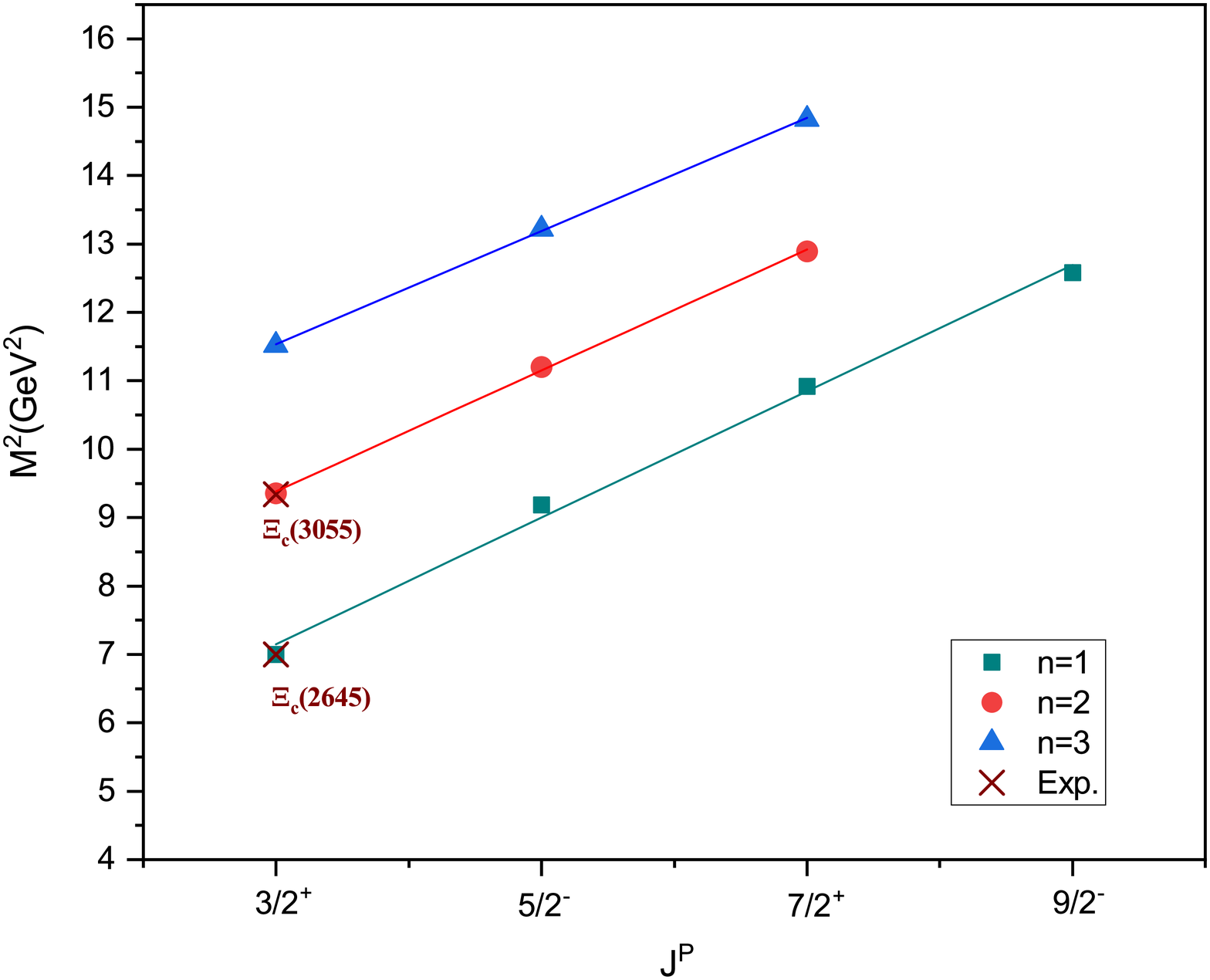}
	\caption{\label{fig:7}{Regge trajectory in the ($J,M^{2}$) plane for $\Xi_{c}^{'}$ baryon for unnatural parity states}}
\end{figure}

\begin{figure}
	\hspace{-2.5cm}
	\includegraphics[scale=0.28]{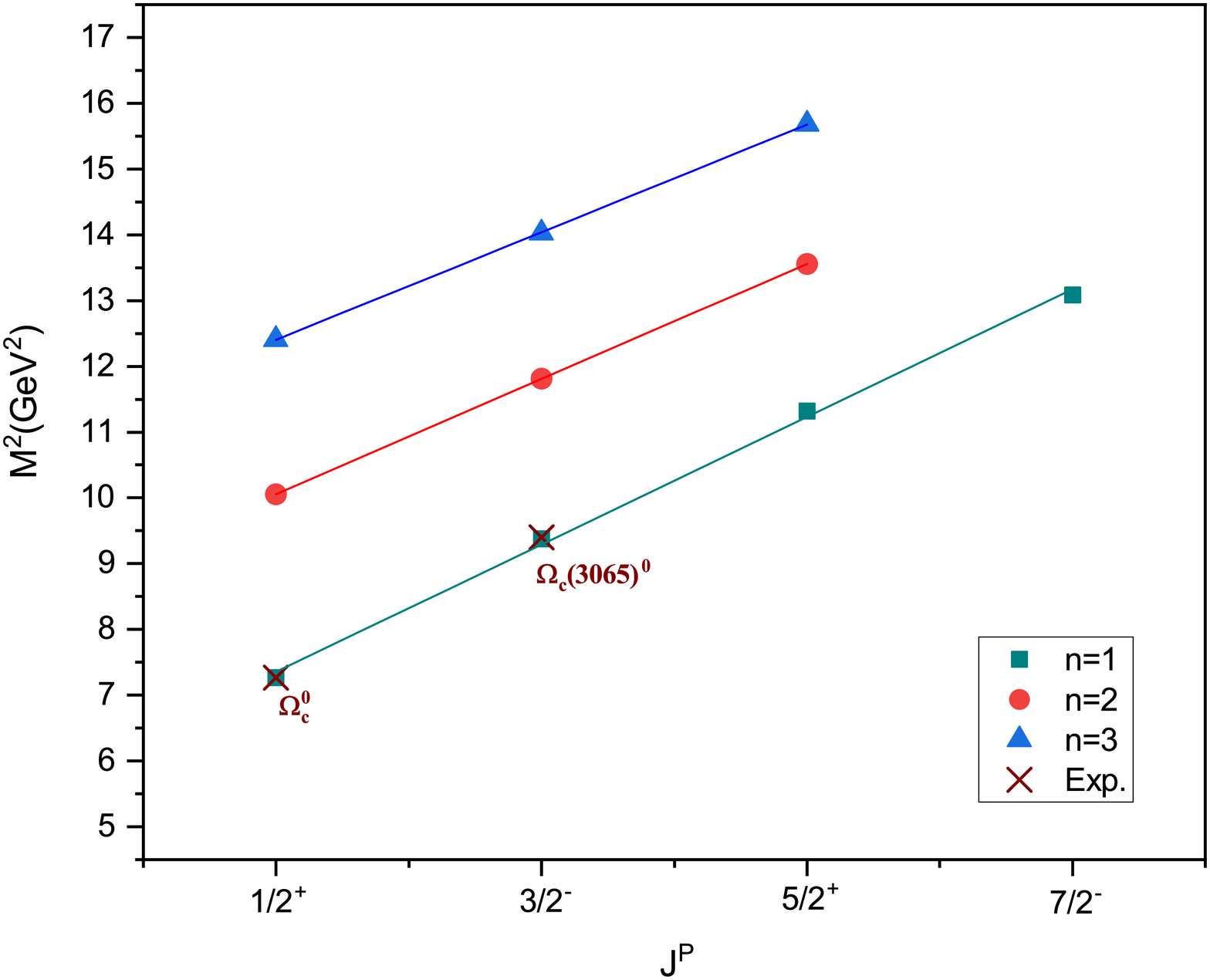}
	\caption{\label{fig:8}{Regge trajectory in the ($J,M^{2}$) plane for $\Omega_{c}$ baryon for natural parity states}}
\end{figure}

\begin{figure}
	\hspace{-2.5cm}
	\includegraphics[scale=0.28]{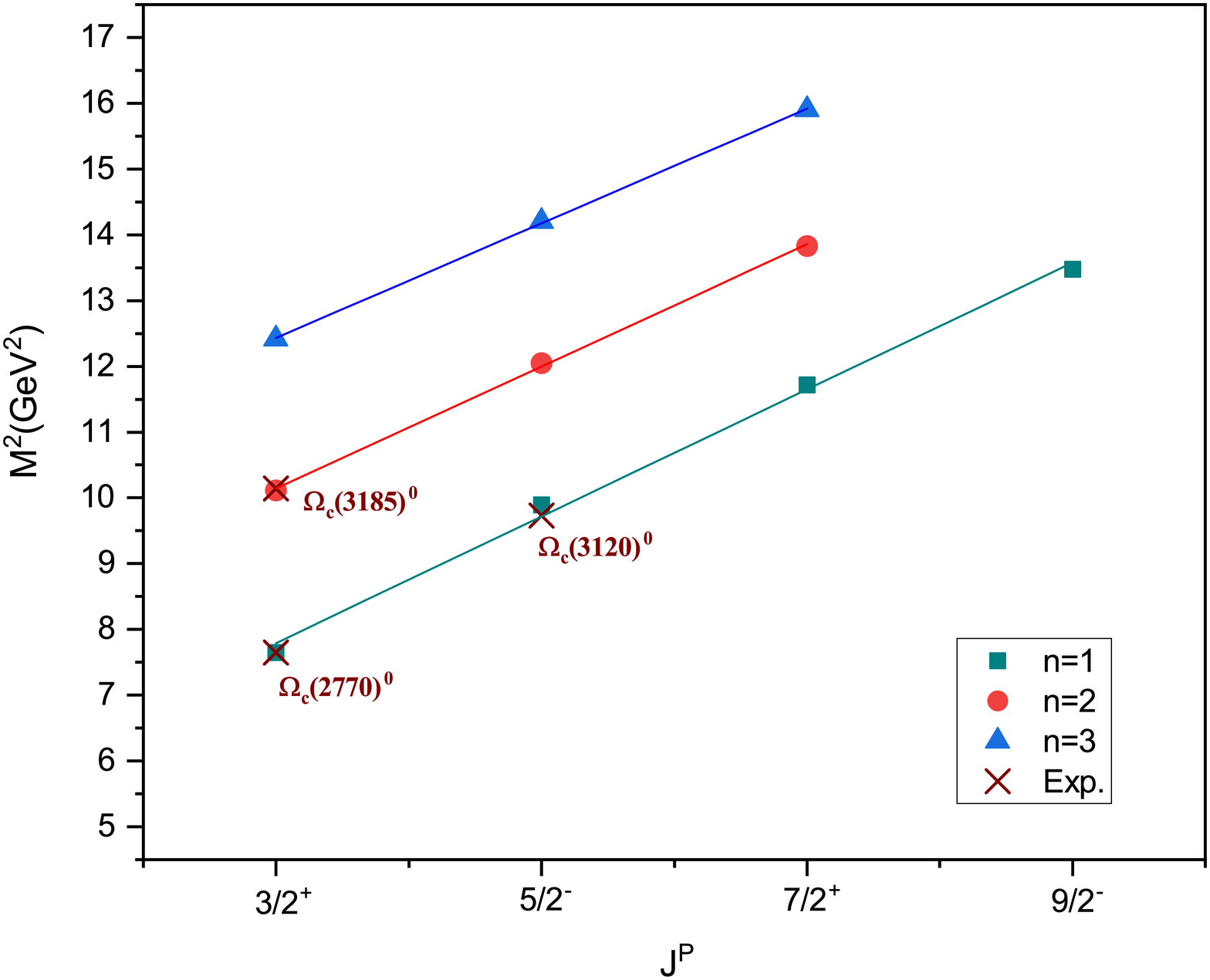}
	\caption{\label{fig:9}{Regge trajectory in the ($J,M^{2}$) plane for $\Omega_{c}$ baryon for unnatural parity states}}
\end{figure}

For $\Sigma_c$, $\Xi'_c$, and $\Omega_c$ baryons, states belonging to the 1P wave have not yet been established experimentally. So, we can't find the value of  $\sigma$ directly from experimental data, as we did for $\Lambda_{c}$ and $\Xi_{c}$ baryons. To find string tension for $\Sigma_c$ and $\Xi'_c$ baryons, authors in Ref. \cite{PhysRevD.101.034016} rely on assumption that the orbital trajectory of $\Lambda_{c}$ and $\Sigma_c$ baryons as well as    $\Xi_c$ and $\Xi'_c$ baryons are parallel, leading to similar string tensions between them. But as shown in Table \ref{tab:table8}, this assumption is not supported by the analysis of mass spectra of singly charmed baryons in relativistic quark potential model \cite{ebert2011, chen2018}. In this model, as the diquark's mass increases, we see that the Regge slope in the (L, $ (\bar{M}-m_{c})^{2} $) plane, or string tension, also increases. 

Within the singly charmed baryonic family, all systems have an identical heavy component, which is a charm quark, but the light diquark has different spin(0 or 1) or quark combinations with which the mass of diquark varies. Hence, the string tension of these systems should be functional of the mass of the diquark only. For simplicity, we assume that string tension is proportional to some power, q, of the mass of diquarks, 
\begin{equation}
	\sigma\propto m_{d}^{q}  .
\end{equation}
A recent study by Song et al. on doubly heavy baryons  uses similar type of power-law assumption for the string tensions of heavy-light hadrons \cite{song2022}.
Taking ratio of string tensions of $ \Xi_{c} $ and $ \Lambda_{c} $, we obtain,
\begin{equation}
	\label{eq:3.1}
	\frac{\sigma_{\Xi_{c}}}{\sigma_{\Lambda_{c}}}=\left(\frac{m_{d_{[d,s]}}}{m_{d_{[u,d]}}}\right)^q.
\end{equation}
By applying the above relation, we first fix $ q=0.661 $.  Then we find  $\sigma$ for ${\Sigma_c}$ ,  ${\Xi_{c}'}$  and  ${\Omega_c} $ baryons, using ratios 
\begin{equation}
	\label{eq:3.2}
	\frac{\sigma_{\Sigma_{c}}}{\sigma_{\Lambda_{c}}}=\left(\frac{m_{d_{\{u,u\}}}}{m_{d_{[u,d]}}}\right)^q,
\end{equation}
\begin{equation}
	\label{eq:3.3}
	\frac{\sigma_{\Xi_{c}'}}{\sigma_{\Lambda_{c}}}=\left(\frac{m_{d_{\{d,s\}}}}{m_{d_{[u,d]}}}\right)^q,
\end{equation}and
\begin{equation}
	\label{eq:3.4}
	\frac{\sigma_{\Omega_{c}}}{\sigma_{\Lambda_{c}}}=\left(\frac{m_{d_{\{s,s\}}}}{m_{d_{[u,d]}}}\right)^q,
\end{equation}
as $\sigma_{\Sigma_c}=1.666 $ GeV$^2$,  $ \sigma_{\Xi_{c}'}= 1.856$ GeV$^2$ and  $ \sigma_{\Omega_c}=2.026 $ GeV$^2$.
Other parameters, $ m_c $, $ v_1 $, $ v_2 $, $\alpha$ and $ b $ fixed for $\Lambda_{c}$ and $\Xi_{c}$, are taken as the inputs.
These parameters are used to determine the masses of  $\Sigma_c$, $\Xi'_{c}$ and $ \Omega_{c} $ baryonic states, which are given in Tables \ref{tab:table4}-\ref{tab:table6}.
From the assumption that string tension depends on q$ ^{th} $ power of the mass of diquark, we have successfully reproduced experimentally detected states of $\Sigma_c$, $\Xi'_{c}$ and $ \Omega_{c} $ baryons.

\section{Results and Discussion}
This section discusses our results for the mass spectra of single charmed baryons. The ground state and excited state masses for $ \Lambda_{c} $, $ \Xi_{c} $, $ \Sigma_{c} $, $ \Xi_{c}' $, and  $ \Omega_{c} $ baryons  are shown in Tables \ref{tab:table2}-\ref{tab:table6}, respectively. These tables provide the baryon quantum numbers $ (n, L, J, j) $, and corresponding baryonic states \textit{$|nL, J^P\rangle$ }  in the first two columns, together with our predicted masses, experimental masses, and prediction from other models in the remaining columns. 
{Further, we have constructed the Regge trajectories for these baryons in the ($J,M^{2}$) plane with natural and unnatural parity states as shown in Figs \ref{fig:2}-\ref{fig:9}. It is found that the experimental masses which are available nicely fit to them. Also it can be seen that these trajectories are almost linear, parallel and equidistant.} We are now attempting to relate the experimentally observed states of singly charmed baryons to our model predictions for the following baryons:

\begin{enumerate}
	
\item{\textit{$\Lambda_{c}$ baryon}}: Our results for mass spectra of $ \Lambda_{c} $ baryon is presented in Table \ref{tab:table2}. The $ 1S $, $ 1P $, and $ 1D  $ states of $ \Lambda_{c} $ baryon are already well established in an experiment.  Masses of these states are well reproduced in our model. PDG lists the highest state $ \Lambda_{c} (2940) $ baryon with  $2939.6^{+1.3}_{-1.5}$ MeV mass. It's favored spin-parity in PDG is $ J^{P}=\frac{3}{2}^{-} $, but it is not certain. The measured mass of  $ \Lambda_{c}(2940) $ baryon is near to our predicted masses for $|2P, 1/2^-\rangle$ and $|2P, 3/2^-\rangle$ states with a slightly higher mass difference of 50 MeV and 61.5 MeV respectively. The predictions in Refs. \cite{ebert2011,chen2018, Yu2022} also shows nearly a same mass difference from the experimental value. Thus, the $ \Lambda_{c}(2940) $ can be assigned to one of 2P state with $ J^{P}=\frac{1}{2}^{-}$ or $ J^{P}=\frac{3}{2}^{-}$.

\item{\textit{$ \Xi_{c} $  and $ \Xi_{c}' $ Baryon}}:                                                                                                      The predicted masses of $ \Xi_{c} $  and $ \Xi_{c}' $ baryon which belongs to anti-triplet and sextet representation, are given in Table \ref{tab:table3} and \ref{tab:table5}, respectively. Experimentally it is not possible to distinguish between excited states of $ \Xi_{c} $  and $ \Xi_{c}' $ baryons. Hence, PDG simply lists them as  $ \Xi_{c} $. 
As shown in Table \ref{tab:table3}, many excited states of $ \Xi_{c} $  and $ \Xi_{c}' $  baryons have been reported experimentally. Due to the small mass difference between these states, it is highly challenging to assign spin-parity to them.
Recently  Belle concluded that  $ \Xi_{c}(2970)$ state belongs to $J^{P}=\frac{1}{2}^{+}$ state with the zero spin of the light-quark degrees of freedom \cite{Bellecas_c2970}. Our theoretical prediction for $|2S, 1/2^+\rangle$ state of the $ \Xi_{c}$  baryon with scalar diquark differs only by 3.4 MeV from the measured mass of the $ \Xi_{c}(2970)$ state. As a result, our calculation agrees very well with Belle's   $J^{P}=\frac{1}{2}^{+}$ assignment for the $ \Xi_{c}(2970)$  state. Moreover, LHCb discovered the $ \Xi_{c}(2965)^{0}$ state \cite{LHCb2020Cas_c}, which lies vary close to the previously observed  $ \Xi_{c}(2970)$ state. More research is needed to determine whether $ \Xi_{c}(2965)^{0}$ is the isospin partner of $ \Xi_{c}(2970)^{+}$ or a distinct state of  $\Xi_c$ baryon. 
Our predictions for the masses of 2S states with $J^{P}=\frac{1}{2}^{+}$ and $\frac{3}{2}^{+}$  of the $\Xi_{c}'$ baryon differ from the experimental mass of $ \Xi_{c}(3055) $ by only 6.6 MeV and 3 MeV, respectively, as shown in Table \ref{tab:table5}. The mass of the $ \Xi_{c}(3055) $ state is also very close to our prediction for the $ \Xi_{c} $ baryon's second orbital excitation (1D)  with $J^{P}=\frac{3}{2}^{+}$. As a result, $ \Xi_c(3055) $ can be interpreted as one of the radial excitations (2S) of $\Xi_c'$ baryon with $J^{P}=\frac{1}{2}^{+}$ or $\frac{3}{2}^{+}$, or it may belong to the 1D state of $\Xi_c$ baryon with $J^{P}=\frac{3}{2}^{+}$. Only through future experiments will it be possible to identify the appropriate assignment.
The $ \Xi_{c}(3080) $ with mass 3079.9 MeV is listed in PDG with a status of three stars. This mass is only 13.4 MeV larger than the model prediction for the 1D state of  $ \Xi_{c}$ baryon with $J^{P}=\frac{5}{2}^{+}$. So, $ \Xi_{c}(3080) $ is good candidate for second orbital excitation(1D) of $ \Xi_{c}$ baryon and we assign $J^{P}=\frac{5}{2}^{+}$ to this state. 
The  $ \Xi_{c}(3123) $ was first observed by BaBar Collaboration \cite{BABARCac_c}. The evidence for this state is quite weak as Belle \cite{BelleCAs_c} didn't find any signal for this state, and PDG lists it with a status of one star only. In our work $|1D, 5/2^+\rangle$ state of $ \Xi_{c}$  baryon and $|1D, 1/2^+\rangle$ state of $ \Xi_{c}'$  baryon lies at 3093.3 MeV and 3157.1 MeV, respectively. Masses of these two states lie relatively closer to the measured mass of $ \Xi_{c}(3123) $ with a deviation of 29.6 MeV and 34.2 MeV, respectively.  Because the $|1D, 5/2^+\rangle$ state of $ \Xi_{c}$  baryon is assigned to the $ \Xi_{c}(3080) $ in our work, the only possibility  for the $ \Xi_{c}(3123) $ is that it is the 1D state of the $ \Xi_{c}' $ baryon  with  $J^{P}=\frac{1}{2}^{+}$.
Finally, the $ \Xi_{c}(2923)$ and $ \Xi_{c}(2938)$ states are interpreted as the first orbital(1P) excitations of $ \Xi_{c}'$ baryon with $J^{P}=\frac{1}{2}^{-}$ and $\frac{3}{2}^{-}$, respectively.



\item{\textit{$\Sigma_{c}$ baryon}}: The masses of $ \Sigma_{c} $ baryonic states as predicted by the RFT model are summarized in Table \ref{tab:table4} with available experimental masses and comparison with other theoretical models. So far only one excited state of $ \Sigma_{c} $ baryon, namely $ \Sigma_{c}(2800) $, has been discovered with mass $ 2801 {}^{+4}_{-6} $ MeV. Its spin and parity have not been identified yet. Our predicted mass 2800.1 MeV for  $|1P, 3/2^-\rangle $ state is in excellent agreement with the experimental mass. Thus  $ \Sigma_{c}(2800) $ is a good candidate for 1P state with $ J^{P}=\frac{3}{2}^{-}$.
Apart from this state many other states belonging to 1P-wave in the vicinity of  $ \Sigma_{c}(2800) $ state, are predicted. These are the states most likely to be detected first in the experiment. This prediction will definitely be helpful for future experiments to detect these unobserved states. The states belonging to 1D and 2S waves are also very likely to be detected in upcoming experiments.  


\item{\textit{$\Omega_{c}$ baryon}}: Recently, five narrow excited states of $ \Omega_{c} $ baryon, namely  $ \Omega_{c}^{0} $, $ \Omega_{c}(3050)^{0} $, $ \Omega_{c}(3065)^{0} $, $ \Omega_{c}(3090)^{0} $, and $ \Omega_{c}(3120)^{0} $, have been detected, the spin-parity of which is unknown. Our predicted masses for $ \Omega_{c} $ states, as shown in Table \ref{tab:table6}, help to determine possible quantum numbers of these experimentally detected states. From the RFT model, we propose that all of these newly observed states of $\Omega_{c}^{0} $  baryon belong to the 1P wave. The experimentally measured mass of the $\Omega_{c}(3000)^{0} $ state, 3000.41 MeV, is very close to the model prediction 3003.2 MeV for $|1P, 1/2^-\rangle_{j=0} $ state. Hence we assign  $J^{P}=\frac{1}{2}^{-}$ for $\Omega_{c}(3000)^{0} $. The measured mass of the $\Omega_{c}(3065)^{0} $ state is only 2.7 MeV higher from the model prediction for the $|1P, 3/2^-\rangle_{j=1} $ state, therefore we simply give  $J^{P}=\frac{3}{2}^{-}$ to the $\Omega_{c}(3065)^{0} $ state. Our theoretical predictions for $ \Omega_{c}(3090)^{0} $, and $ \Omega_{c}(3120)^{0} $   differ  by 16 MeV and 26 MeV, respectively from their experimental masses. Accordingly, it is acceptable to assign them $J^{P}=\frac{3}{2}^{-}$ and $J^{P}=\frac{5}{2}^{-}$, respectively. The $|1P, 1/2^-\rangle {j=1} $ state with $J^{P}=\frac{1}{2}^{-}$ is eventually identified as the $ \Omega {c}(3050)^{0} $, however its predicted mass is underestimated by 40 MeV. These five spin-parity assignments are also supported by Ref. \cite{karliner2017,jia2021}.
\textbf{Additionally, we predict the spin and parity of  two newly discovered states $\Omega_{c}(3185)^{0}$ and $\Omega_{c}(3327)^{0}$. 
Our theoretical prediction for $|2S, 3/2^+\rangle$ state is only 5 Mev less than the experimental mass of $\Omega_{c}(3185)^{0}$ state. So, we assign $ J^{P}={\frac{3}{2}^{+}}$ to  $\Omega_{c}(3185)^{0}$ state.
Finally, since the experimentally measured mass of the $\Omega_{c}(3327)^{0}$ state only differs by a maximum of 6 MeV from our prediction for the $|1D, 3/2^+\rangle$  state, we assign the  $ J^{P}={\frac{3}{2}^{+}}$  to this state.}

\end{enumerate}	

Further, we compare our results with existing theoretical predictions made using the quark-diquark picture \cite{ebert2011,chen2017,chen2015}  and the three-body picture \cite{rbrt2008,shah2016} of baryon.
Ebert et al. have studied the mass spectra of heavy baryons up to quite high excitations($ L=5 $, $ n_{r}=5 $) using a QCD-motivated relativistic quark potential model with a quark-diquark picture of baryons \cite{ebert2011}. We put their results in Tables \ref{tab:table2}-\ref{tab:table6} as a key reference for comparison with our results. 
For $ \Lambda_{c} $  and $ \Xi_{c} $ baryons our predictions are in excellent agreement with this reference. Up to $ 3S $, $ 3P $, $ 3D $, $ 3F $, $ 2G $, and $ 1H $ states of $ \Lambda_{c} $ baryon, our predictions differ by a maximum of 32.7 MeV only and as we move to some more radially excited states, this difference slowly increases. For the $ \Xi_{c} $  baryon, our prediction for states up to  $ 4S, 4P, 4D, 3F, 2G, $ and $ 1H $ deviates from ref. \cite{ebert2011} at most by 61 MeV only. 
Our calculated masses for  $ \Sigma_{c} $, $ \Xi_{c}' $, and $ \Omega_{c} $ baryons are also consistent enough with this  ref. \cite{ebert2011} for states belonging to $ S $-wave and $ P $-wave, and $ D $-wave. 
In ref. \cite{chen2017}, the non-relativistic constituent quark model has been employed and a quark-diquark picture has been considered to investigate the mass spectra of  $ \Lambda_{c} $, $ \Xi_{c}, $ $\Sigma_{c}$, and $\Xi_{c}'$ baryons. Our predictions and the results of this model are in accordance, although the difference rises for higher orbital and radial excited states.
Chen et al. have investigated mass spectra of $ \Lambda_{c} $ and $ \Xi_{c} $ baryons in quark-diquark framework \cite{chen2015} with relativistic flux tube model. 
The spin-orbit interaction term in this work differs from our model since the Thomas-precession term is also included in our work.  Though, its contribution is relatively small as it is inversely proportional to the square of the diquark's mass.
The masses of the ground state and few excited states of singly charmed baryons were also studied in a three-body picture of baryon with quark model in ref. \cite{rbrt2008} and its findings are consistent with those of our model.
We also compare our results for $ \Lambda_{c} $, $ \Sigma_{c} $, and $ \Omega_{c} $ baryons with ref. \cite{shah2016} in which hyper central constituent quark model is employed with three body picture. This model's prediction for states belonging to a single orbital excitation is such that the state with higher J lies below the state with lower J, which is one of its limitations. However, no such inconsistency is seen in our model.

\textbf{Since, mass of some the experimental candidates are quite close to  more than one  of our calculated masses,  we have assigned more than one possible  spin-parity quantum number to these experimental states.  The most prominent way for eliminating some of these possibilities is to calculate the decay widths of these states. In Ref. \cite{E. Abreu2007} the authors linearly fitted the decay width ($\Gamma$) of the light mesons with the string  length (RMS) for maximal J states (where J = L + S1 + S2 , and  S1 and S2 are the spins of quark and antiquark, respectively.) as,  
\begin{equation}
	\Gamma = \gamma(RMS - r_{0}) \pm \Delta \Gamma, 
\end{equation}
where $\gamma$ = 0.05 $\pm$ 0.01 GeV$^{2}$ and $r_{0}$ = 1.4 $\pm$ 0.6 GeV$^{-1}$. This linear relation is then extrapolated to glueballs. In our work, the total flux tube length dependent on two quantum numbers, n and L (see Eq. (10)). Now to check whether such linear relation between decay width and string length exists or not for singly charmed baryons, experimental decay widths of at least three maximal J states is required. In the future, when sufficient experimental data on decay width will be available, the relation between decay width and string length can be studied to assign a spin-parity quantum number for singly charmed baryons.}

\section{Conclusion}
In this work, we used a mass formula derived from a relativistic flux tube model to investigate mass spectra of singly charmed baryons in a heavy quark-light diquark framework. Due to the strong coupling between two light quarks, it is less probable that at the low-energy region, baryon with excited diquark could be detected. So, we only considered states in which the diquark in the ground state excites orbitally or radially with respect to charm quark, as these states are more likely to be detected first in the experiments. The spin-dependent interactions were included in the j-j coupling scheme. The experimentally well-known states of singly charmed baryons can be well reproduced and their $J^{P}$ values have also been confirmed by considering them as a system of heavy quark and light diquark connected by a mass-loaded flux tube.
For low-lying orbital and radial excited states, our outcomes are consistent with many theoretical models, but for higher excited states, we observe a variety of model-dependent differences.
Our predicted mass spectra help us to assign the possible spin-parity of experimentally detected states such as  $\Sigma_{c}(2800)$, $\Xi_{c}(2923)$, $\Xi_{c}(2930)$, $\Xi_{c}(2970)$, $\Xi_{c}(3055)$, $\Xi_{c}(3080)$, and $\Xi_{c}(3123)$, as well as all five states of  $\Omega_{c}$ baryon, including  $\Omega_{c}(3000)$, $\Omega_{c}(3050)$, $\Omega_{c}(3065)$, $\Omega_{c}(3090)$, and $\Omega_{c}(3119)$. We have also predicted many unobserved states of singly charmed baryons which have a good potential to be detected first in the experiment. These predictions can be used as reference data for upcoming experimental searches like CMS, LHCb, Belle II, BESIII \cite{BESIII2022}, and  PANDA \cite{PANDA}  for heavy hadron physics.

\textbf{Despite the fact that we employed a quark-diquark picture, the possibility of a three-body picture of a singly charmed baryon may not be removed. The three body relativistic flux tube model has also been obtained in Ref.\cite{brambilla1995} from Wilson area law in QCD. But, for three body system it is very difficult to obtain Regge relation between mass and angular momentum. Therefore, utilising the three-body relativistic flux tube model to describe the mass spectra of singly charmed baryons remains an open problem. Additionally, due to this, we were unable to find connection between the two-body and three-body flux tube model, and it is still not clear how the quark-diquark relativistic flux tube develops from the three-body relativistic flux tubes. 
} 

\textbf{After the successful determination of mass spectra of singly charmed baryons, we will extend this model to study singly bottom, doubly and triply charmed and bottom baryons. }

\section{Acknowledgment}
Ms. Pooja Jakhad acknowledges the financial assistance by the Council of Scientific \& Industrial Research (CSIR) under the JRF-FELLOWSHIP scheme with file no. 09/1007(13321)/2022-EMR-I. 

\section{Appendix}
\begin{widetext}
Here, we go into detail on how to obtain mass-splitting operators that are involved in spin-dependent interactions for singly charmed baryons with vector diquark only. The following outline only three orbitally excited states for demonstration purpose.

\begin{enumerate}
\item {\textit{The S-wave}}: For this state L=0. In spin-dependent interactions, only spin-spin contact hyperfine interaction survive. Expectation value of  $ \mathbf{S_d}.\mathbf{S_c} $, in both L-S and J-j coupling is same.
\item {\textit{The P-wave}}:We have three angular momentum vectors $ \mathbf{S_d} $, $ \mathbf{S_c} $ and $  \mathbf{L} $. In L-S coupling scheme,   $ \mathbf{S_d} $ and $ \mathbf{S_c} $ first couple to give  $\mathbf{S} $.  The simultaneous eigenstate of  $\mathbf{S}$ and it's third component ${S_{3}}$ can be constructed from uncoupled states $ |S_{d}, S_{d_{3}}\rangle $ and $ |S_{c}, S_{c_{3}}\rangle $ as \cite{brink1968}  
\begin{equation}\label{A1}
	|S_{d}S_{c};S\ S_{3}\rangle = \displaystyle\sum_{S_{d_{3}}S_{c_{3}}} C_{S_{d_{3}}S_{c_{3}}S_{3}}^{S_{d}\ S_{c}\ S}\ |S_{d}S_{d_{3}}\rangle |S_{c}S_{c_{3}}\rangle.
\end{equation}
Then, $\mathbf{S}$ combine with $\mathbf{L}$ to generate total angular momentum  $\mathbf{J}$. The simultaneous eigenstate of $\mathbf{J}$, $ J_{3} $ and $\mathbf{S}$ can be formed by  $ |S_{d}S_{c};S\ S_{3}\rangle $ and uncoupled state $ |L\ L_{3}\rangle $ as
\begin{equation}\label{A2}
	\begin{split}
		|(S_{d}S_{c})SL;J\ J_{3}\rangle &=\displaystyle\sum_{\scriptscriptstyle {S_{3}L_{3}}}	C_{S_{3}L_{3}J_{3}}^{S\ L\ J}\ |S_{d}S_{c};S\ S_{3}\rangle|L\ L_{3}\rangle\\
		&= \displaystyle\sum_{\scriptscriptstyle {S_{d_{3}}S_{c_{3}}L_{3}}S_{3}} C_{S_{d_{3}}S_{c_{3}}S_{3}}^{S_{d}\ S_{c}\ S}\ 
		C_{S_{3}L_{3}J_{3}}^{S\ L\ J}\ |S_{d}S_{d_{3}}\rangle |S_{c}S_{c_{3}}\rangle  |L\ L_{3}\rangle,
	\end{split}
\end{equation}

where, 	${ C_{S_{d_{3}}S_{c_{3}}S_{3}}^{S_{d}\ S_{c}\ S}}$ and $ C_{S_{3}L_{3}J_{3}}^{S\ L\ J}$ are Clebsch-Gordan coefficients. $ S_{d_{3}} $, $ S_{c_{3}} $, $ L_{3} $ and $ J_{3} $ denotes third component of respective angular momentum. For simplicity, we abbreviate basis $ |(S_{d}S_{c})SL;J\ J_{3}\rangle $  as  $ |^{2S+1}L_{J};J_{3}\rangle $, and the product states $|S_{d}S_{d_{3}}\rangle |S_{c}S_{c_{3}}\rangle  |L\ L_{3}\rangle$ as $|S_{d_{3}},S_{c_{3}},L_{3}\rangle$ for fixed value of $ S_{d} $, $ S_{c} $ and $ L $.
Then, the L-S coupling basis states can be constructed as a linear combination of $|S_{d_{3}},S_{c_{3}},L_{3}\rangle$ states using 
\begin{equation}\label{A3}
	|^{2S+1}L_{J};J_{3}\rangle = \displaystyle\sum_{\scriptscriptstyle {S_{d_{3}}S_{c_{3}}L_{3}}S_{3}} C_{S_{d_{3}}S_{c_{3}}S_{3}}^{S_{d}\ S_{c}\ S}\ C_{S_{3}L_{3}J_{3}}^{S\ L\ J}\ |S_{d_{3}},S_{c_{3}},L_{3}\rangle.  
\end{equation}
Finally, utilizing the above relation, the L-S basis are constructed for the P-wave which are listed below \cite{karliner2015}:

\begin{equation} \label{A4}
	|^{2}P_{1/2};1/2\rangle =-\frac{\sqrt{2}}{3}|0,-\frac{1}{2},1\rangle+\frac{\sqrt{2}}{3}|1,-\frac{1}{2},0\rangle+\frac{2}{3}|-1,\frac{1}{2},1\rangle-\frac{1}{3}|0,\frac{1}{2},0\rangle ,
\end{equation}
\begin{equation}\label{A5}
	|^{4}P_{1/2};1/2\rangle =\frac{1}{3}|0,-\frac{1}{2},1\rangle-\frac{1}{3}|1,-\frac{1}{2},0\rangle+\frac{1}{3 \sqrt{2}}|-1,\frac{1}{2},1\rangle-\frac{\sqrt{2}}{3}|0,\frac{1}{2},0\rangle+\frac{1}{\sqrt{2}}|1,\frac{1}{2},-1\rangle ,
\end{equation}
\begin{equation}\label{A6}
	|^{2}P_{3/2};3/2\rangle =\sqrt{\frac{2}{3}}|1,-\frac{1}{2},1\rangle-\frac{1}{\sqrt{3}}|0,\frac{1}{2},1\rangle ,
\end{equation}
\begin{equation}\label{A7}
	|^{4}P_{3/2};3/2\rangle =-\sqrt{\frac{2}{15}}|1,-\frac{1}{2},1\rangle-\frac{2}{\sqrt{15}}|0,\frac{1}{2},1\rangle+\sqrt{\frac{3}{5}}|1,\frac{1}{2},0\rangle \  \text{and}
\end{equation}
\begin{equation}\label{A8}
	|^{4}P_{5/2};5/2\rangle =|1,\frac{1}{2},1\rangle.
\end{equation}
Now we define the operators involved in spin-dependent interactions. The operator $ \mathbf{L}.\mathbf{S_i}$ in  terms of raising and lowering operator is given by
\begin{equation}
	\mathbf{L.S_{i}}=\frac{1}{2}\left[L_{+}S_{i-}+L_{-}S_{i+}\right]+L_{3}S_{i3},
\end{equation}
where i = l or h. The operator engaged in tensor interaction term can be simplified to  \cite{karliner2017}
\begin{equation}
	\hat{B}=\frac{-3}{(2L-1)(2L+3)}\left[(\mathbf{L.S_{d}})(\mathbf{L.S_{c}})+(\mathbf{L.S_{c}})(\mathbf{L.S_{d}})-\frac{2}{3}L(L+1)(\mathbf{S_{d}.S_{c}})\right].
\end{equation}
Squaring the identity $ \mathbf{S}=\mathbf{S_{d}}+\mathbf{S_{c}} $ allows one to calculate the expectation value of the operator $ \mathbf{S_{d}.S_{c}} $ as 
\begin{equation}
\langle\mathbf{S_d}.\mathbf{S_c}\rangle=\frac{1}{2}[S(S+1)-S_d(S_d+1)-S_c(S_c+1)].
\end{equation}

With these operators in hand, we determine its expectation value in L-S  basis  [$^{2}P_{J} ,^{4}P_{J}$] for different possible values of J as listed below:\\
For J=1/2,
\begin{equation}\label{A9}
	\text{$\langle\mathbf{L}.\mathbf{S_d}\rangle$=}\left[
	\begin{array}{cc}
		-\frac{4}{3} & -\frac{\sqrt{2}}{3} \\
		-\frac{\sqrt{2}}{3} & -\frac{5}{3} \\
	\end{array}
	\right],\ \   \text{$\langle\mathbf{L}.\mathbf{S_c}\rangle$=}\left[
	\begin{array}{cc}
		\frac{1}{3} & \frac{\sqrt{2}}{3} \\
		\frac{\sqrt{2}}{3} & -\frac{5}{6} \\
	\end{array}
 	\right],\ \   \text{$\langle\hat{B}\rangle$=}\left[
	\begin{array}{cc}
		0 & \frac{1}{\sqrt{2}} \\
		\frac{1}{\sqrt{2}} & -1 \\
	\end{array}
	\right],\ \ 
	\text{$\langle\mathbf{S_d}.\mathbf{S_c}\rangle$=}\left[
	\begin{array}{cc}
		-1 & 0 \\
		0 & \frac{1}{2} \\
	\end{array}
	\right].
\end{equation}
For J=3/2,
\begin{equation}\label{A10}
\text{$\langle\mathbf{L}.\mathbf{S_d}\rangle$=}\left[
\begin{array}{cc}
	\frac{2}{3} & -\frac{\sqrt{5}}{3} \\
	-\frac{\sqrt{5}}{3} & -\frac{2}{3} \\
\end{array}
\right],\ \ \text{$\langle\mathbf{L}.\mathbf{S_c}\rangle$=}\left[
\begin{array}{cc}
	-\frac{1}{6} & \frac{\sqrt{5}}{3} \\
	\frac{\sqrt{5}}{3} & -\frac{1}{3} \\
\end{array}
\right],\ \ \text{$\langle\hat{B}\rangle$=}\left[
\begin{array}{cc}
	0 & -\frac{1}{2 \sqrt{5}} \\
	-\frac{1}{2 \sqrt{5}} & \frac{4}{5} \\
\end{array}
\right],\ \ \text{$\langle\mathbf{S_d}.\mathbf{S_c}\rangle$=}\left[
\begin{array}{cc}
	-1 & 0 \\
	0 & \frac{1}{2} \\
\end{array}
\right]
\end{equation}
For J=5/2,
\begin{equation}\label{A11}
\text{$\langle\mathbf{L}.\mathbf{S_d}\rangle$ =1}	,\ \text{$\langle\mathbf{L}.\mathbf{S_c}\rangle$ = }\frac{1}{2},\ \text{$\langle\hat{B}\rangle$=}-\frac{1}{5},\ \text{$\langle\mathbf{S_d}.\mathbf{S_c}\rangle$=}\frac{1}{2}.
\end{equation}
As $ m_{c}\gg m_{d}$, the term proportional to $ L.S_{d} $ dominates over other terms involved in spin-dependent interactions. $ L.S_{d} $ matrix is diagonal in $|J, j \rangle $ basis in j-j coupling. Therefore, it is reasonable to employ $|J, j \rangle$ basis where the dominant interaction is diagonal and other interactions are treated perturbatively. For each eigenvalue $\lambda$  of $ L.S_{d} $ with specific J, we find the corresponding eigenvector, which forms $|J, j \rangle $ basis as listed below:
\begin{equation}\label{A12}
\text{ $\lambda $ = }-2\text{ : $|$J=}\frac{1}{2}\text{, j=0}\text{$\rangle$ = }\frac{1}{\sqrt{3}}\text{ $|$$^{2}P_{1/2}$}\text{$\rangle$ + }\sqrt{\frac{2}{3}}\text{ $|$$^{4}P_{1/2}$}\text{$\rangle$ },
\end{equation}
\begin{equation}\label{A13}
\text{ $\lambda $ = }-1\text{ : $|$J=}\frac{1}{2}\text{, j=1}\text{$\rangle$ = }-\sqrt{\frac{2}{3}}\text{ $|$$^{2}P_{1/2}$}\text{$\rangle$ + }\frac{1}{\sqrt{3}}\text{ $|$$^{4}P_{1/2}$}\text{$\rangle$ },
\end{equation}
\begin{equation}\label{A14}
\text{ $\lambda $ = }-1\text{ : $|$J=}\frac{3}{2}\text{, j=1}\text{$\rangle$ = }\frac{1}{\sqrt{6}}\text{ $|$$^{2}P_{3/2}$}\text{$\rangle$ + }\sqrt{\frac{5}{6}}\text{ $|$$^{4}P_{3/2}$}\text{$\rangle$ },
\end{equation}
\begin{equation}\label{A15}
\text{ $\lambda $ = }1\text{ : $|$J=}\frac{3}{2}\text{, j=2}\text{$\rangle$ = }-\sqrt{\frac{5}{6}}\text{ $|$$^{2}P_{3/2}$}\text{$\rangle$ + }\frac{1}{\sqrt{6}}\text{ $|$$^{4}P_{3/2}$} \text{$\rangle$ },
\end{equation}
\begin{equation}\label{A16}
	\text{$|$J=}\frac{5}{2}\text{, j=2}\text{$\rangle$ = $|$$^{4}P_{5/2}$}\text{$\rangle$  }.
\end{equation}
This gives baryonic states in heavy quark limit. Now, we determine the expectation value for each mass splitting operator in $|J, j \rangle $ basis and list the results in table \ref{tab:table7}.

\item {\textit{The D-wave}}:
For D-wave, the same process as for P-wave is used to produce the mass-splitting operators. We first use  Eq. (\ref{A3})  to construct L-S basis and the outcomes are 
\begin{equation}
|^{4}D_{1/2};1/2\rangle =-\sqrt{\frac{2}{5}}|-1,-\frac{1}{2},2\rangle+\frac{1}{\sqrt{5}}|0,-\frac{1}{2},1\rangle-\frac{1}{\sqrt{15}}|1,-\frac{1}{2},0\rangle+\frac{1}{\sqrt{10}}|-1,\frac{1}{2},1\rangle-\sqrt{\frac{2}{15}}|0,\frac{1}{2},0\rangle+\frac{1}{\sqrt{10}}|1,\frac{1}{2},-1\rangle,
\end{equation}
\begin{equation}
|^{2}D_{3/2};3/2\rangle =-\frac{2}{\sqrt{15}}|0,-\frac{1}{2},2\rangle+\sqrt{\frac{2}{15}}|1,-\frac{1}{2},1\rangle+2 \sqrt{\frac{2}{15}}|-1,\frac{1}{2},2\rangle-\frac{1}{\sqrt{15}}|0,\frac{1}{2},1\rangle,
\end{equation}
\begin{equation}
|^{4}D_{3/2};3/2\rangle =\frac{2}{\sqrt{15}}|0,-\frac{1}{2},2\rangle-\sqrt{\frac{2}{15}}|1,-\frac{1}{2},1\rangle+\sqrt{\frac{2}{15}}|-1,\frac{1}{2},2\rangle-\frac{2}{\sqrt{15}}|0,\frac{1}{2},1\rangle+\frac{1}{\sqrt{5}}|1,\frac{1}{2},0\rangle,
\end{equation}
\begin{equation}
|^{2}D_{5/2};5/2\rangle =\sqrt{\frac{2}{3}}|1,-\frac{1}{2},2\rangle-\frac{1}{\sqrt{3}}|0,\frac{1}{2},2\rangle,
\end{equation}
\begin{equation}
|^{4}D_{5/2};5/2\rangle =-\frac{2}{\sqrt{21}}|1,-\frac{1}{2},2\rangle-2 \sqrt{\frac{2}{21}}|0,\frac{1}{2},2\rangle+\sqrt{\frac{3}{7}}|1,\frac{1}{2},1\rangle\  \text{and}
\end{equation}
\begin{equation}
|^{4}D_{7/2};7/2\rangle =|1,\frac{1}{2},2\rangle.
\end{equation}
Secondly, the expectation of mass splitting operators for specific J, in L-S  basis  [$^{2}D_{J} ,^{4}D_{J}$,] are computed and listed below:\\
For J=1/2,
\begin{equation}
	\text{$\langle\mathbf{L}.\mathbf{S_d}\rangle$=}-3,\ \text{$\langle\mathbf{L}.\mathbf{S_c}\rangle$=}-\frac{3}{2},\ \text{$\langle\hat{B}\rangle$=}-1,\ \text{$\langle\mathbf{S_d}.\mathbf{S_c}\rangle$=}\frac{1}{2}.
\end{equation}
For J=3/2,
\begin{equation}
\text{$\langle\mathbf{L}.\mathbf{S_d}\rangle$=}\left[
\begin{array}{cc}
	-2 & -1 \\
	-1 & -2 \\
\end{array}
\right],\ \ \text{$\langle\mathbf{L}.\mathbf{S_c}\rangle$=}\left[
\begin{array}{cc}
	\frac{1}{2} & 1 \\
	1 & -1 \\
\end{array}
\right],\ \ \text{$\langle\hat{B}\rangle$=}\left[
\begin{array}{cc}
	0 & \frac{1}{2} \\
	\frac{1}{2} & 0 \\
\end{array}
\right],\ \ \text{$\langle\mathbf{S_d}.\mathbf{S_c}\rangle$=}\left[
\begin{array}{cc}
	-1 & 0 \\
	0 & \frac{1}{2} \\
\end{array}
\right]	.
\end{equation}
For J=5/2,
\begin{equation}
\text{$\langle\mathbf{L}.\mathbf{S_d}\rangle$=}\left[
\begin{array}{cc}
	\frac{4}{3} & -\frac{\sqrt{14}}{3} \\
	-\frac{\sqrt{14}}{3} & -\frac{1}{3} \\
\end{array}
\right],\ \ \text{$\langle\mathbf{L}.\mathbf{S_c}\rangle$=}\left[
\begin{array}{cc}
	-\frac{1}{3} & \frac{\sqrt{14}}{3} \\
	\frac{\sqrt{14}}{3} & -\frac{1}{6} \\
\end{array}
\right],\ \ \text{$\langle\hat{B}\rangle$=}\left[
\begin{array}{cc}
	0 & -\frac{1}{\sqrt{14}} \\
	-\frac{1}{\sqrt{14}} & \frac{5}{7} \\
\end{array}
\right],\ \ \text{$\langle\mathbf{S_d}.\mathbf{S_c}\rangle$=}\left[
\begin{array}{cc}
	-1 & 0 \\
	0 & \frac{1}{2} \\
\end{array}
\right].
\end{equation}
For J=7/2,
\begin{equation}
\text{$\langle\mathbf{L}.\mathbf{S_d}\rangle$ =2},\ \text{$\langle\mathbf{L}.\mathbf{S_c}\rangle$ =1},\ \text{$\langle\hat{B}\rangle$=}-\frac{2}{7},\ \text{$\langle\mathbf{S_d}.\mathbf{S_c}\rangle$=}\frac{1}{2}.
\end{equation}
In the third step, eigenvalue $ \lambda $ and eigenvector of $\langle\mathbf{L}.\mathbf{S_d}\rangle$ are determined and $|J, j \rangle $ basis are constructed as a linear combination of L-S basis with coefficients depending on eigenvector of  $\langle\mathbf{L}.\mathbf{S_d}\rangle$.
\begin{equation}
\text{$|$J=}\frac{1}{2}\text{, j=1}\text{$\rangle$ = $|$$^{4}D_{1/2}$}\text{$\rangle$  },
\end{equation}
\begin{equation}
	\text{ $\lambda $ = }-3\text{ : $|$J=}\frac{3}{2}\text{, j=1}\text{$\rangle$ = }\frac{1}{\sqrt{2}}\text{ $|$$^{2}D_{3/2}$}\text{$\rangle$ + }\frac{1}{\sqrt{2}}\text{ $|$$^{4}D_{3/2}$}\text{$\rangle$ },
\end{equation}
\begin{equation}
\text{ $\lambda $ = }-1\text{ : $|$J=}\frac{3}{2}\text{, j=2}\text{$\rangle$ = }-\frac{1}{\sqrt{2}}\text{ $|$$^{2}D_{3/2}$}\text{$\rangle$ + }\frac{1}{\sqrt{2}}\text{ $|$$^{4}D_{3/2}$}\text{$\rangle$ },
\end{equation}
\begin{equation}
\text{ $\lambda $ = }2\text{ : $|$J=}\frac{5}{2}\text{, j=2}\text{$\rangle$ = }-\frac{\sqrt{7}}{3}\text{ $|$$^{2}D_{5/2}$}\text{$\rangle$ + }\frac{\sqrt{2}}{3}\text{ $|$$^{4}D_{5/2}$}\text{$\rangle$ },
\end{equation}

\begin{equation}
	\text{ $\lambda $ = }-1\text{ : $|$J=}\frac{5}{2}\text{, j=3}\text{$\rangle$ = }\frac{\sqrt{2}}{3}\text{ $|$$^{2}D_{5/2}$}\text{$\rangle$ + }\frac{\sqrt{7}}{3}\text{ $|$$^{4}D_{5/2}$}\text{$\rangle$ },
\end{equation}
\begin{equation}
\text{$|$J=}\frac{7}{2}\text{, j=3}\text{$\rangle$ = $|$$^{4}D_{7/2}$}\text{$\rangle$ . }	
\end{equation}
Finally, we find the expectation value of mass splitting operators in $|J, j \rangle $ basis and collect our results in Table \ref{tab:table7}. 
\end{enumerate}

\end{widetext}

\end{document}